\documentclass[journal=jacsat,manuscript=article]{achemso}

\setkeys{acs}{articletitle = true}
\SectionsOn
\SectionNumbersOn
\usepackage[version=3]{mhchem} 
\usepackage{ulem}
\usepackage{chemfig}
\usepackage{subfig}
\usepackage{graphicx}
\usepackage{hyperref}
\usepackage{bm}
\usepackage{color}
\usepackage{tabularray}
\usepackage{oplotsymbl}

\usepackage{amssymb}

\DeclareMathOperator{\vtheta}{\boldsymbol{\theta}}
\DeclareMathOperator{\Lrho}{{\cal L}_{\rho}}

\DeclareMathOperator{\Loss}{{\cal L}}
\definecolor{Mercury}{rgb}{0.898,0.898,0.898}
\author{Carlos A. Martins Junior}
\affiliation{University of São Paulo, Department of Materials Physics and Mechanics, Institute of Physics, Rua do Matão 1371, São Paulo, 05508-090, Brazil}

\author{Daniela A. Damasceno}
\affiliation{University of São Paulo, Department of Mechatronics and Mechanical Systems Engineering, Polytechnic School, Av. Professor Mello Moraes, 2231, São Paulo, 05315-970, Brazil}

\author{Keat Yung Hue}
\affiliation{Imperial College London, Department of Chemical Engineering, South Kensington Campus, London, SW7 2AZ, UK}
\alsoaffiliation{PETRONAS Research Sdn. Bhd., Lot 3288 \& 3289, Off Jalan Ayer Itam, Kawasan Institusi Bangi, 43000 Kajang, Selangor, Malaysia}

\author{Caetano R. Miranda}
\affiliation{University of São Paulo, Department of Materials Physics and Mechanics, Institute of Physics, Rua do Matão 1371, São Paulo, 05508-090, Brazil}

\author{Erich A. Müller}
\affiliation{Imperial College London, Department of Chemical Engineering, South Kensington Campus, London, SW7 2AZ, UK}

\author{Rodrigo A. Vargas-Hern\'{a}ndez}
\email{vargashr@mcmaster.ca}
\affiliation[Mc]{Department of Chemistry and Chemical Biology, McMaster University, Hamilton, ON, Canada}
\alsoaffiliation[BIMR]{Brockhouse Institute for Materials Research, McMaster University, Hamilton, ON, Canada}

\title{Scalable Bayesian Optimization for High-Dimensional Coarse-Grained Model Parameterization} 

\abbreviations{Coarse-Grained Force Field,Bayesian Optimization,Pebax}
\keywords{American Chemical Society, \LaTeX}

\begin{document}



\newpage
\begin{abstract}
Coarse-grained (CG) force field models are extensively utilised in material simulations due to their scalability. Traditionally, these models are parameterized using hybrid strategies that integrate top-down and bottom-up approaches; however, this combination restricts the capacity to jointly optimize all parameters. While Bayesian Optimization (BO) has been explored as an alternative search strategy for identifying optimal parameters, its application has traditionally been limited to low-dimensional problems. This has contributed to the perception that BO is unsuitable for more realistic CG models, which often involve a large number of parameters. In this study, we challenge this assumption by successfully extending BO to optimize a high-dimensional CG model. Specifically, we show that a 41-parameter CG model of Pebax-1657, a copolymer composed of alternating polyamide and polyether segments, can be effectively parameterized using BO, resulting in a model that accurately reproduces key physical properties of its atomistic counterpart. Our optimization framework simultaneously targets density, radius of gyration, and glass transition temperature. It achieves convergence in fewer than 600 iterations, resulting in a CG model that shows consistent improvements across all three properties.
\end{abstract}

\section{Introduction}
Molecular dynamics (MD) is a well-known computational technique widely used to study physical, chemical, and transport properties at the nanoscale \cite{hollingsworth2018molecular,karplus2002molecular}. By employing molecular models such as all-atom or united-atom representations, MD has allowed researchers to predict the behavior of complex systems under a wide range of thermodynamic conditions \cite{Sadeghi2020,Kostritskii2021,Ding2021}. All-atom models offer the highest level of molecular detail by representing each atom individually, making them ideal for capturing phenomena governed by precise atomic-scale interactions \cite{rahman2018saft, avendano2011saft}. In contrast, united-atom models consolidate specific groups of atoms into a single interaction site, effectively simplifying the system and reducing computational demands. In both approaches, accuracy relies on the choice of force fields, which describe interatomic interactions and enable the calculation of physical properties such as diffusion coefficients, thermal conductivity, elastic constants, and more. However, some phenomena extend beyond the molecular scale, requiring a mesoscale perspective to capture their dynamics effectively. This necessity highlights the importance of coarse-grained (CG) models, as emphasized in several reviews over recent years \cite{joshi2021review,noid2023perspective,brini2013systematic}.

CG models simplify molecular representations by grouping atoms or functional groups into beads. This simplification inevitably results in a loss of chemical resolution, but when carefully designed, CG models can retain key structural and thermodynamic properties of the system \cite{potter2019assessing}. Two main approaches guide the development of CG models: i) bottom-up and ii) top-down. Bottom-up frameworks derive CG parameters by mapping detailed atomistic or quantum-level information to a coarser scale, often using techniques such as iterative Boltzmann inversion \cite{reith2003deriving} or force matching \cite{izvekov2004effective}. This family of methodologies emphasizes accuracy in reproducing molecular-level properties but lacks transferability across diverse thermodynamic conditions \cite{potter2019assessing}. Conversely, the top-down approaches leverage experimental macroscopic data to optimize directly the CG parameters, aiming for broader applicability. Advanced top-down methods include the MARTINI force field \cite{souza2021martini}, widely used for proteins, and the Statistical Associating Fluid Theory (SAFT-$\gamma$ Mie) equation of state (EoS), broadly applied to polymers \cite{fayaz2022coarse}, greenhouse gases \cite{avendano2013saft}, mixtures \cite{lobanova2016saft}, and other systems \cite{papaioannou2016application}.

Hybrid approaches combine the strengths of top-down and bottom-up methodologies to develop CG models that are computationally efficient and capable of accurately representing molecular interactions. For example, Rahman et al. \cite{rahman2018saft} integrated the SAFT-$\gamma$ Mie with atomistic simulations to model linear alkanes. The SAFT-$\gamma$ Mie EoS represents molecules as chains of tangentially bonded spherical segments interacting via the Mie potential—a generalization of the Lennard-Jones potential that allows for independent tuning of repulsive and attractive forces. In the top-down phase, SAFT-$\gamma$ Mie is employed to derive nonbonded interaction parameters by fitting experimental thermophysical data, such as densities, vapor pressures, and phase equilibria, ensuring that CG models align with macroscopic observations. The bottom-up phase complements this by refining bonded interactions through atomistic simulations, maintaining consistency with structural features at the molecular level. 

As emphasized in previous works \cite{rahman2018saft}, the importance of accurately capturing both intra- and intermolecular interactions is critical for obtaining reliable predictions of thermodynamic properties such as vapor-liquid equilibrium and densities. Similarly, Fayaz-Torshizi et al. \cite{fayaz2022coarse} applied these methodologies to polymers, demonstrating that incorporating detailed interactions significantly enhances the model's ability to predict structural and dynamic behavior. By mapping atomistic configurations to CG beads, the models ensure that the bond lengths and angles closely reflect the equilibrium states observed experimentally. The harmonic potentials governing bond stretching and angular bending were calibrated using distributions of bond lengths and angles derived from atomistic simulations. These distributions were typically fitted with weighted Gaussian functions, ensuring consistency with key structural features such as end-to-end distances and radii of gyration.
Although effective, this sequential optimization process may occur in isolated stages, providing no assurance of achieving a global solution, \emph{i.e.}, the optimal set of CG parameters. The decoupled nature of the optimization may lead to suboptimal CG models. 

Machine learning (ML)-based CG models have also emerged as an alternative parameterization of CG models \cite{durumeric2023machine}. However, despite significant progress, challenges remain, such as the need for extensive datasets and ensuring model transferability across diverse systems \cite{durumeric2023machine}.
Other alternative schemes dependent on ML models use Bayesian optimization (BO), a search algorithm, to identify the optimal parameters for CG models. 
BO is a powerful tool for design optimization, widely applied across diverse fields such as robotics, environmental monitoring, and experimental design \cite{BO_review}. Within computational chemistry, BO has demonstrated its effectiveness in various applications, including screening chemical compounds \cite{ueno2016combo,jalem2018bayesian,ju2017designing}, minimizing the energy of the Ising model \cite{tamura2018bayesian}, and optimizing laser pulses for molecular control \cite{deng2020bayesian}. Furthermore, BO has been used successfully to calibrate functional density models \cite{vargas2020}, physical models for the $cis$–$trans$ photoisomerization of retinal in rhodopsin \cite{Vargas2021,singh2024}, and to design potential energy surfaces for reactive molecular systems \cite{vargas2019}.

BO provides a systematic and efficient approach for exploring high-dimensional parameter spaces~\cite{kandasamy15:HDBO, wang16:HDBO, nayebi19a:HDBO, eriksson19:HDBO, eriksson21a:HDBO, papenmeier22:HDBO, ziomek23a:HDBO, hvarfner24a:HDBO}, making it particularly well-suited for CG models of nanoscale materials, where traditional methods often struggle. Its application to compact CG models has been explored in Refs. \cite{Weeratunge2023, sestito2020}, where the optimization process involved extracting physical properties from MD simulations and incorporating them into an objective function, typically defined as the sum of relative errors compared to an atomistic model or within a multi-objective optimization framework.
The resulting CG models demonstrated strong agreement with their atomistic counterparts, underscoring BO’s potential for improving CG model development. However, these studies were restricted to CG models with relatively few parameters \cite{cordina2023, sestito2020, Weeratunge2023} and did not fully account for all inter- and intra-molecular interactions during optimization, limiting the robustness of the resulting models.

Here, we demonstrate that BO can efficiently optimize CG models with a significantly larger number of parameters while integrating both bottom-up and top-down approaches. In our copolymer case study, the parameter search space exceeds 40 dimensions, showcasing BO’s scalability and its ability to navigate and refine intricate molecular interaction landscapes, ultimately enhancing the accuracy and versatility of CG models.

The paper is structured as follows: Section \ref{sec:sec_cg_model} introduces the CG model framework for copolymers, using Pebax as a representative example. Section \ref{sec:bo} provides a brief overview of the BO algorithm for optimizing CG models, with an emphasis on the physical properties used for model calibration. Finally, Section \ref{sec:results} presents the results and discussion of the findings.

\section{Coarse-Grained Models for Copolymers}\label{sec:sec_cg_model}
\subsection{Pebax-1657}
\begin{figure}[H]
\centering
\includegraphics[width=\textwidth]{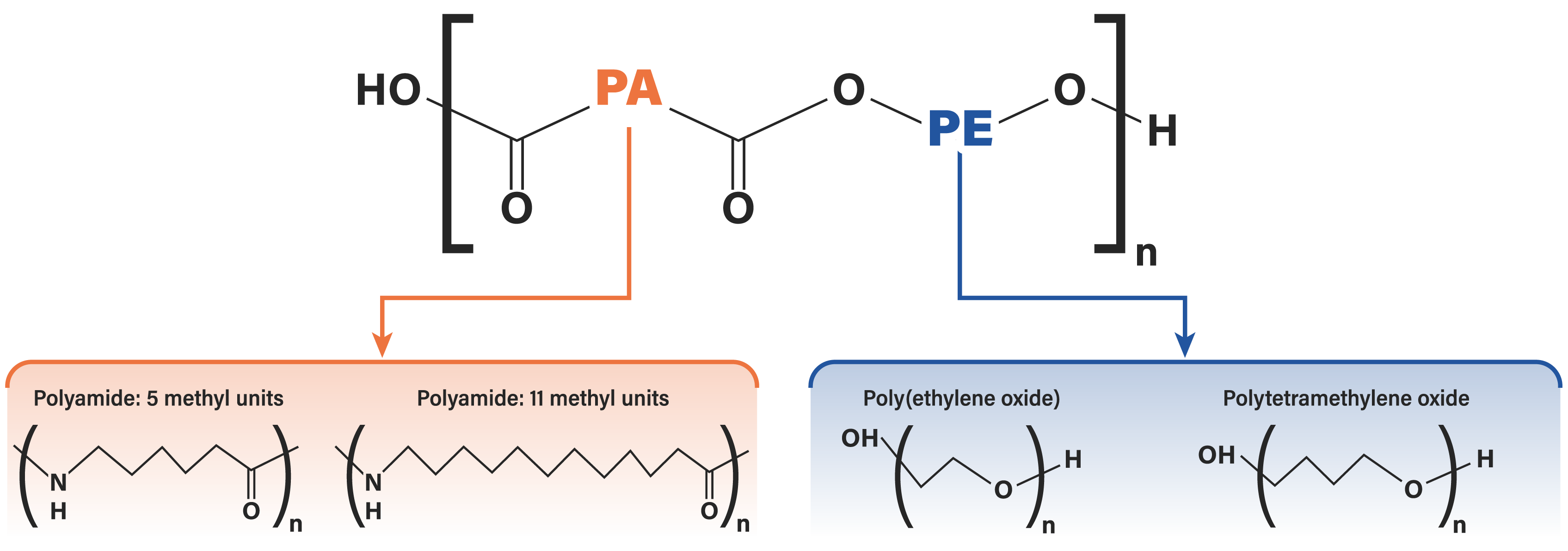}
\caption{Chemical structure of Pebax and its polyamide (PA) and polyether (PE) segments.}
\label{fig:Pebax}
\end{figure}

\begin{figure}[H]
\centering
\includegraphics[width=\textwidth]{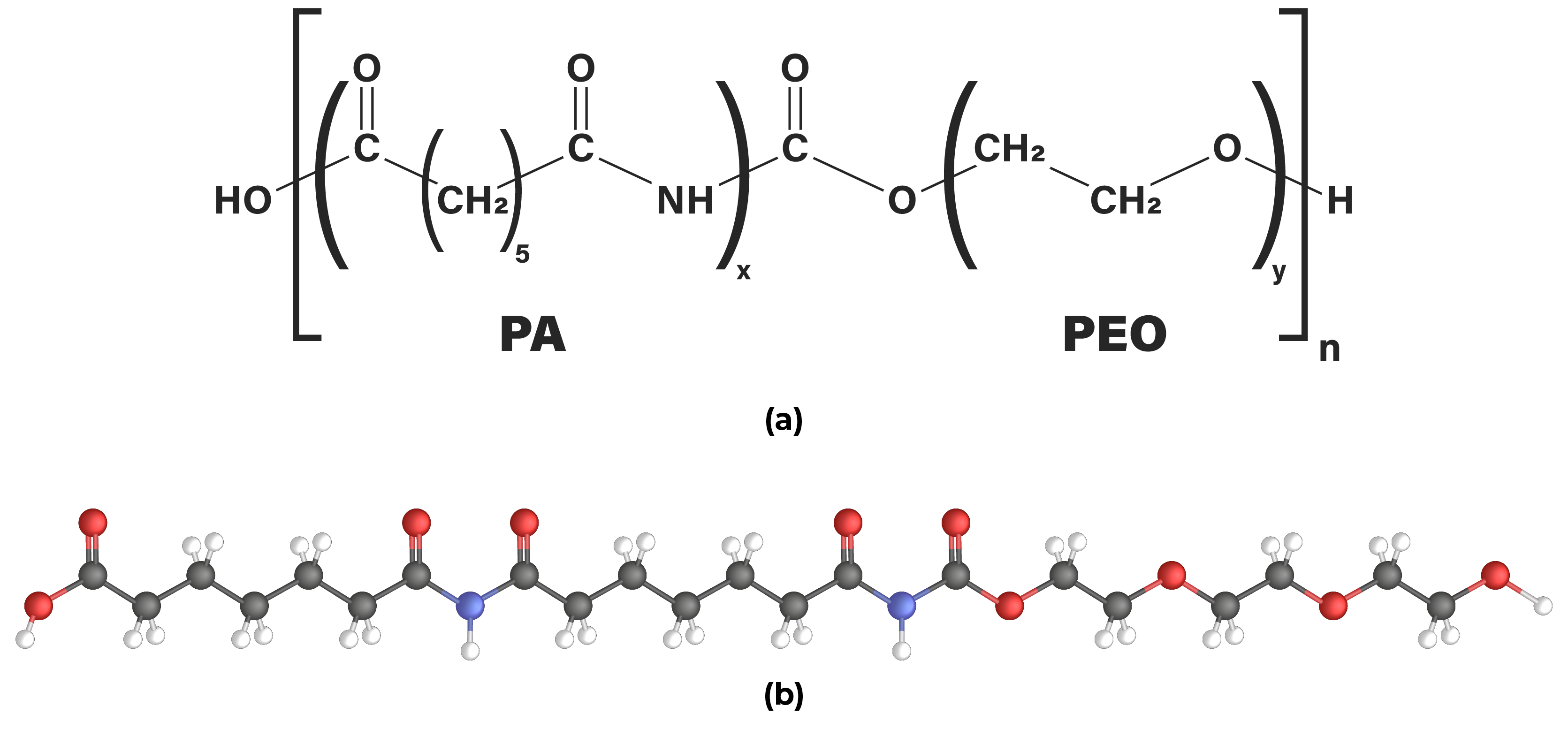}
\caption{Panel a), Chemical structure of Pebax-1657 composed of  PA6 and PEO segments.
Panel b), AA model of the Pebax-1657 chain with a composition of 40\% PA6 and 60\% PEO, corresponding to n = 1, x = 2, and y = 3.}
\label{fig:Pebax_1657}
\end{figure}

Pebax copolymers, composed of alternating polyamide and polyether segments, are widely used in membrane technologies due to their tunable mechanical and transport properties. Pebax consists of polyamide (PA) and polyether (PE) segments, as shown in Fig. \ref{fig:Pebax}. The PA segment, typically polyamide 6 (PA6) or polyamide 12 (PA12), provides mechanical strength and is covalently linked to the PE segment via ester groups \cite{embaye2021poly}. The PE component, which can be poly(ethylene oxide) (PEO) or poly(tetramethylene oxide) (PTMO) \cite{embaye2021poly}, enhances gas transport and separation.
As a case study, we focus on Pebax-1657, a specific grade composed of PEO and PA segments, as illustrated in Fig. \ref{fig:Pebax_1657}a. The repeat unit of the Pebax-1657 chain consists of 40\% PA6 and 60\% PEO \cite{khalilinejad2015preparation,meshkat2018mixed,bernardoenhancing}, with each chain composed of one repeat unit only, as illustrated in Fig.~\ref{fig:Pebax_1657}b. The PA-to-PEO ratio plays a crucial role in balancing mechanical flexibility and gas permeability, making Pebax-1657 particularly suitable for membrane applications such as \ce{CO2} capture and gas separation \cite{embaye2021poly,salestan2021experimental}. Additionally, the incorporation of nanomaterials—including \ce{MoS2} nanosheets \cite{liu2019characterization}, nonionic surfactants \cite{bernardoenhancing}, zeolitic imidazolate frameworks-8 (ZIF-8) particles \cite{li2020enhanced}, and Azo@MOF-199 \cite{xin2023preparation}—has been shown to enhance its physical and transport properties further \cite{salestan2021experimental, jiang2021facile}. Given its technological potential, Pebax-1657 is the polymer of choice in this work.

The atomistic (AA) representation of the Pebax-1657 polymer was constructed using the Material Exploration and Design Analysis (MedeA) simulation software (version 3.5) \cite{medea}. The system consists of 50 polymer chains arranged in an amorphous configuration, generated using the Amorphous Builder module. The dimensions of the initial simulation box were set to 32.6 Å × 32.6 Å × 32.6 Å (length × width × height), yielding a density of 1.136 g/cm\textsuperscript{3}. 

\subsection{Pebax CG Model}
\begin{figure}[H]
\centering
\includegraphics[width=\textwidth]{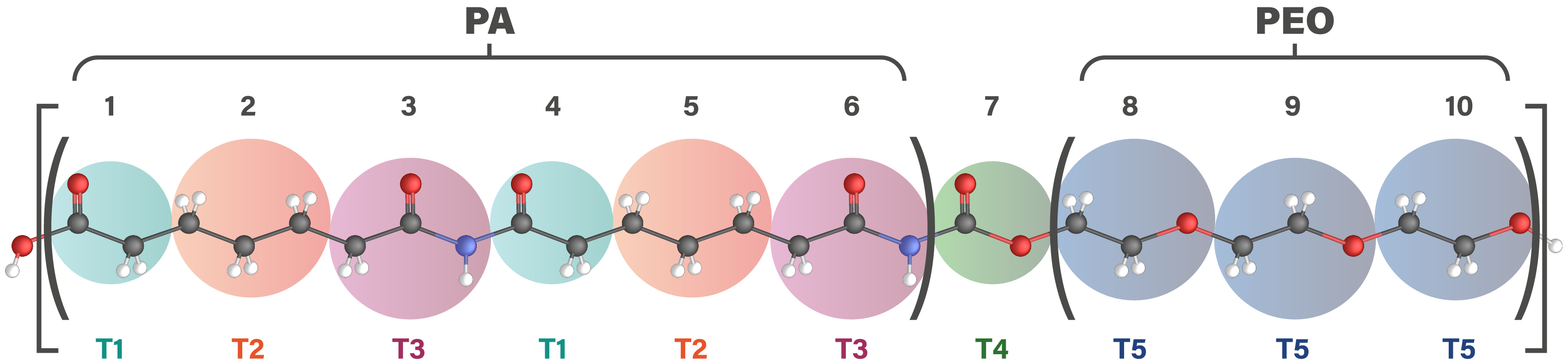}
\caption{CG model of Pebax-1657 chain considering five distinct beads. The numbers 1 to 10 clearly visualize the structural organization and the total number of beads.}
\label{fig:CG_Pebax}
\end{figure}

Given the technological relevance of Pebax-1657, an accurate coarse-grained (CG) model is essential for studying its structural and transport properties at larger scales. The CG model, shown in Fig. \ref{fig:CG_Pebax}, was developed based on available data from small oligomers, assuming the transferability of parameters. The atom-to-bead mapping followed a protocol to determine the intermolecular parameters of the CG model using the SAFT-$\gamma$ Mie group contribution method. The model classifies the beads into five types: T1, T2, T3, T4, and T5, reflecting their chemical structure and functional roles within the polymer. The T1, T2, and T3 beads represent the polyamide segment, while T5 corresponds to the polyether segment. The T4 bead serves as a linker between the PA and the PEO regions. 

\begin{figure}[H]
\centering
\includegraphics[width=\textwidth]{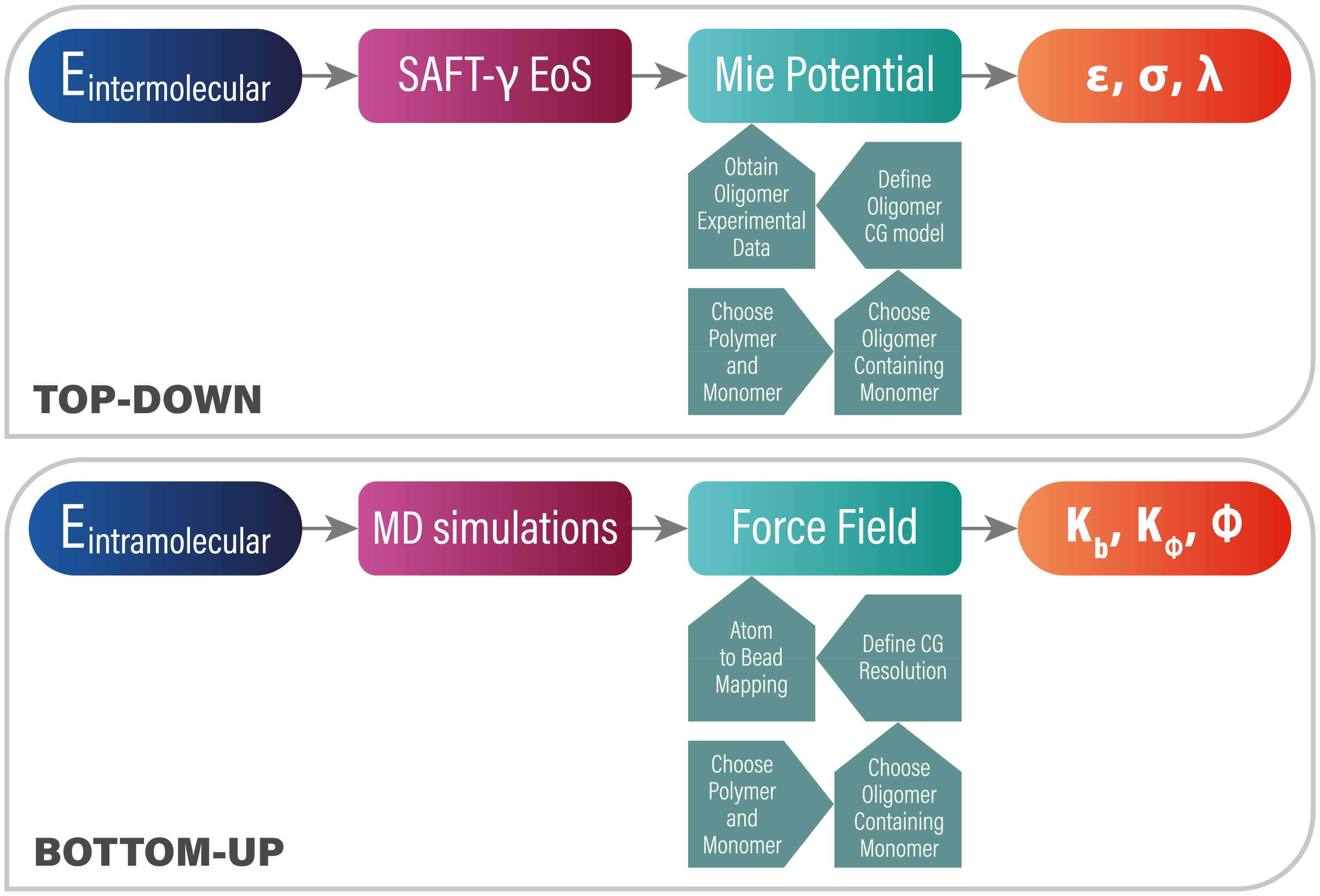}
\caption{Hybrid strategy combining top-down and bottom-up methodologies to develop CG models.}
\label{fig:Top_bottom}
\end{figure}

For this proposed CG model, a hybrid strategy combining top-down and bottom-up methodologies was employed to derive both intermolecular and intramolecular FF parameters. Fig.~\ref{fig:Top_bottom} summarizes the standard procedure to obtain the non-bonded intermolecular parameters [$\epsilon$, $\sigma$, $\lambda$] using the SAFT-$\gamma$ Mie group contribution method, where $\epsilon$, $\sigma$, and $\lambda$ represent the depth of the potential well, bead diameter, and Mie potential parameter that controls repulsion and attraction contributions, respectively. 
All intramolecular interactions were modeled using harmonic potentials for bond stretching and angle bending \cite{rahman2018saft}, and these parameters [$K_b$, $K_\phi$, $\phi$] were derived through fully atomistic simulations using MD simulations based on the PCFF+ potential~\cite{sun1998}. 

In the SAFT formalism, the SAFT-$\gamma$ Mie framework determines non-bonded intermolecular parameters by minimizing an objective function ($\Loss$) that compares experimental data with predictions from the equation of state (EoS) for target properties such as saturated liquid densities and vapor pressures \cite{fayaz2022coarse}. This approach enables the extraction of top-down interaction parameters based on macroscopic bulk properties, making it a valuable tool for parameter estimation \cite{muller2014force}.
A widely used approach for optimizing CG models is Square Gradient Theory (SGT) combined with the SAFT-VR Mie EoS. For example, SGTPy, a Python-based implementation of SGT, provides a practical framework for such optimizations \cite{mejia2021sgtpy}. 

Rather than following the standard procedure to obtain the Mie potential parameters, we used Bottled SAFT \cite{ervik2016bottled}.
Bottled SAFT is a web application that provides SAFT-$\gamma$ Mie parameters for a wide range of molecular fluids, allowing researchers to search for a specific molecule or upload a structure file to generate a set of parameters. Bottled SAFT is freely available at \url{http://www.bottledsaft.org} and is widely recognized for its accessibility and utility. For beads T1, T3, and T4, the parameters were obtained, considering heptanoic acid (CAS 111-14-8), N-methylacetamide (CAS 79-16-3), and pentyl pentanoate (CAS 2173-56-0), respectively. While Bottled-SAFT parameters derived from n segments are typically used to construct coarse-grained models for homonuclear molecules, where the CG model includes all $n$ segments, several studies have adopted an alternative approach to build the CG model. Notably, in the CG modeling of polystyrene \cite{Guadalupe2017}, a single representative segment for toluene from the n segments was selected to construct the CG model. In line with this precedent, we believe that selecting a representative segment from each relevant molecule, along with its corresponding bottled-SAFT parameters, offers a more accurate representation for our CG model. Meanwhile, data for bead T2 and the PEO segment (T5) were derived from previous studies of similar chain groups \cite{Richards2022}. For intramolecular FFs, all the steps illustrated in Fig.~\ref{fig:Top_bottom} were performed to derive the bond stretching parameters ($K_b$) and angle bending parameters ([$K_\phi$, $\phi$]).

\subsection{Optimization}
The development of CG models commonly depends on hybrid strategies that combine top-down and bottom-up methodologies, as shown in Fig. \ref{fig:Top_bottom}. In these hybrid approaches, the optimization is divided into two sub-problems: i) the intermolecular and ii) the intramolecular parameters. 
First, the optimal parameters of the intermolecular potential $[\epsilon,\sigma, \lambda]$ are determined by minimizing the error between the reference data and the EoS calculations. Once the intermolecular parameters are optimized, a bottom-up approach follows to tune the intramolecular potential parameters, $[K_{b}, K_{\phi}, \phi]$, for harmonic force fields. 

Although hybrid approaches have been shown to be successful~\cite{rahman2018saft, ervik2016bottled, muller2014force, lafitte2012saft, fayaz2021}, the division of the global optimization problem into separate subproblems can lead to inefficient exploration of the search space. This decomposition often results in suboptimal solutions, as it fails to capture the interdependencies between parameter groups. To address this limitation, an alternative optimization approach is needed: one that explicitly accounts for these interdependencies while remaining computationally feasible, even for high-dimensional parameter spaces and complex physical systems.
BO is often regarded as unsuitable for high-dimensional problems, primarily due to the challenge of constructing accurate surrogate models in large parameter spaces with limited data. This perception is reflected in prior applications to CG model optimization, where the number of parameters is typically limited to $\sim 10$ or fewer~\cite{cordina2023, sestito2020, Weeratunge2023}. Despite these dimensional limitations, BO has been successfully applied to developing CG models by optimizing force fields based on physical properties extracted from molecular dynamics simulations~\cite{Weeratunge2023, sestito2020}. 
While existing applications of BO to CG models have been constrained to relatively low-dimensional problems, recent advances have demonstrated that BO can be effective in high-dimensional settings~\cite{kandasamy15:HDBO, wang16:HDBO, nayebi19a:HDBO, eriksson19:HDBO, eriksson21a:HDBO, papenmeier22:HDBO, ziomek23a:HDBO, hvarfner24a:HDBO}. For example, in Ref.~\cite{hvarfner24a:HDBO}, BO was successfully applied to an optimization setup involving up to $1,000$ variables, showcasing its scalability and adaptability to large, complex search spaces.
These findings suggest that BO could be an effective tool for optimizing high-dimensional CG models and overcoming prior limitations on existing optimization algorithms where all inter- and intra-molecular interactions are not fully accounted by existing optimization frameworks, limiting the robustness of the resulting models.

For copolymers such as Pebax, the three primary physical properties used to assess the CG model's accuracy are (i) density ($\rho$), (ii) radius of gyration ($Rg$), and (iii) glass transition temperature ($Tg$). The proposed optimization framework integrates these three properties into a single objective function,
\begin{eqnarray} 
{\cal L}(\vtheta) = w_\rho {\cal L}_{\rho}(\vtheta) + w_{Rg}{\cal L}_{Rg}(\vtheta) + w_{Tg}{\cal L}_{Tg}(\vtheta), \label{eqn:L_total} 
\end{eqnarray}
where $\mathbf{w} = [w_\rho, w_{Rg}, w_{Tg}]$ are weight coefficients that balance the relative importance of each property. These weights ensure that no single property disproportionately influences the optimization process. We used $w_\rho = 1, w_{Rg} = 1,875$, and $w_{Tg} = 1.5\times10^4$, see Supplementary Information (SI) for more discussion regarding these values. 
The individual loss terms in Eq.~\ref{eqn:L_total} are computed as relative errors between the CG model predictions and the reference data,
\begin{eqnarray} 
\label{eqn:objective_parts} 
{\cal L}_{\rho}(\boldsymbol{\theta}) &=& \sum_{i=1}^{N_\rho} \left( \frac{\rho(T_i, \bm{\theta}) - \hat{\rho}(T_i)}{\hat{\rho}(T_i)} \right)^2, \label{eqn:L_rho}\\ 
{\cal L}_{Rg}(\boldsymbol{\theta}) &=& \sum_{i=1}^{N_{Rg}} \left( \frac{Rg(T_i,\bm{\theta}) - \hat{R}g(T_i)}{\hat{R}g(T_i)} \right)^2, \label{eqn:L_rg} \\
{\cal L}_{Tg}(\boldsymbol{\theta}) &=& \left( \frac{Tg(\bm{\theta}) - \hat{T}g}{\hat{T}g} \right)^2, \label{eqn:L_tg}  
\end{eqnarray}
where $T_i$ denotes the temperatures at which MD simulations were performed, while $N_\rho$ and $N_{Rg}$ represent the number of data points used for density and radius of gyration calculations, respectively. These properties play distinct roles in the validation of CG models. The density provides direct experimental comparability and is accurately simulated in MD, the radius of gyration reflects the spatial conformation and compactness of polymer chains, and the glass transition temperature integrates density and structural information, offering insight into thermal behavior. 

For the copolymer Pebax, $\Lrho$ is influenced by the glass temperature. To ensure representative sampling, we selected eight discrete temperatures ($\{T_i\}_{i=1}^{8}$), four below and four above the phase transition observed in the atomistic model. Temperatures near the glass transition were excluded, as the density-temperature relationship in this region deviates from linearity, instead exhibiting a hyperbolic trend~\cite{PATRONE2016}. 
We used density measurements at each of the eight temperatures to fit two linear curves, each generating $7,500$ interpolated points; see Fig. S.I in the SI. This approach provided a denser set of target points on either side of the phase transition, refining the density-temperature relationship for optimization. Initially, we included points near the phase transition in the fits. However, we found that excluding them led to improved results. The glass transition temperature was then determined as the intersection of these two fitted curves, following the methodology outlined by Patrone \textit{et al.}\cite{PATRONE2016}. 
Further numerical details are provided in the SI.

For the radius of gyration component, $\Loss_{Rg}$, we also ensured reliable sampling by simulating a single Pebax chain instead of a fully populated polymer box. This approach mitigated the risk of the MD simulations becoming confined to a limited region of phase space, which could otherwise lead to inaccurate property estimates. By isolating a single chain, we enhanced sampling across phase space. The radius of gyration was then computed at 15 different temperature points. Additional details on the molecular dynamics simulations are provided in the SI.
All target properties used as reference data in $\Loss$ (Eq. ~\ref{eqn:L_total}) were determined using an atomistic model based on the PCFF+ force field, and all MD simulations used the LAMMPS package \cite{LAMMPS} on a 24 × 12th Gen Intel(R) Core(TM) i9-12900K with 64GB of memory. The data supporting the findings of this study are openly available in the following \url{https://github.com/camjjr/bo_cgff}

\subsection{Bayesian Optimization}\label{sec:bo}
Grounded in Bayesian inference, BO systematically balances exploration and exploitation by iteratively updating a surrogate model that navigates the search space efficiently. By leveraging information from previous evaluations, BO aims to identify high-potential regions and direct the search toward the global optimum.
Two key components conform to the core of BO: the surrogate model, which approximates the objective function, and the acquisition function, which guides the search.
We used as a surrogate model the Tree-structured Parzen Estimator (TPE)\cite{tpe2023}. Unlike traditional Gaussian process-based approaches, TPE constructs a non-parametric density estimator that partitions the search space into two regions: (i) a top quantile of observations with favorable objective function values and (ii) a lower quantile containing the remaining observations\cite{tpe2023}. This structure allows TPE to model the search space adaptively, prioritizing promising regions while maintaining diversity in exploration.
Formally, TPE models parameterize the conditional probability distribution as,
\begin{eqnarray}
    p({\bf \vtheta}|y,D)= 
\begin{cases}
    p({\bf \vtheta}|D^w)& y \le y^\gamma\\
    p({\bf \vtheta}|D^b)& y > y^\gamma
\end{cases} \label{eqn:tpe_model}
\end{eqnarray}
where $p({\bf \vtheta}|y,D)$ represents the probability density function of the parameters, and $D$ is the set of observed objective function values $\{y_1,y_2,y_3,\cdots, y_n\}$ obtained from evaluations of Eq.~\ref{eqn:L_total}. The subsets $D^w$ and $D^b$ correspond to the best (b) observations and worst (w), respectively. These groups are dynamically updated at each iteration based on the hyperparameter $\gamma$, which controls the balance between exploration and exploitation.
The probability density function within each group is estimated as follows,
\begin{equation} 
p({\bf \vtheta}|D^i) = \omega_0^i p_0({\bf \vtheta}) + \sum_{n=1}^{N_i} \omega_n^i k_n({\bf \vtheta},{\bf \vtheta_n}|b^i),\label{eqn:tpe_prob} 
\end{equation}
where $\omega^i$ represents the weights assigned to observations, $i$ denotes either the best (b) or worst (w) group, and $N_i$ is the number of observations within the group. The term $p_0$ defines a prior distribution that influences the level of exploration, while $k_n({\bf \vtheta},{\bf \vtheta_n}|b^i)$ is the kernel function used for density estimation, employing a Gaussian kernel for numerical variables and an Aitchison-Aitken kernel for categorical variables. $b^i$ is simply the bandwidth for each group. 
For this study, we used the following acquisition function,
\begin{eqnarray}
    {\cal P}(y \leq y^* | \vtheta, D) = \int_{-\infty} ^{y^*} p(y|\vtheta, D) dy.
\end{eqnarray}
Using Eq.~\ref{eqn:tpe_model}, the PI acquisition function is~\cite{tpe2023},
\begin{equation} 
{\cal P}(y \leq y^* | \vtheta, D) \overset{\text{rank}}{\simeq} r({\bf \vtheta}|D) = \frac{p({\bf \vtheta}|y,D^b)}{p({\bf \vtheta}|y,D^w)}. \label{eqn:acq_fun}
\end{equation}
which is equivalent to the well-established PI acquisition function \cite{bayesoptbook,tpe2023}. 
The TPE model and BO algorithm were implemented using the Optuna library~\cite{optuna_2019}. 

\section{Results}\label{sec:results}
\subsection{CG Optimization}
\begin{figure}[H]
\centering
\includegraphics[scale=0.6]{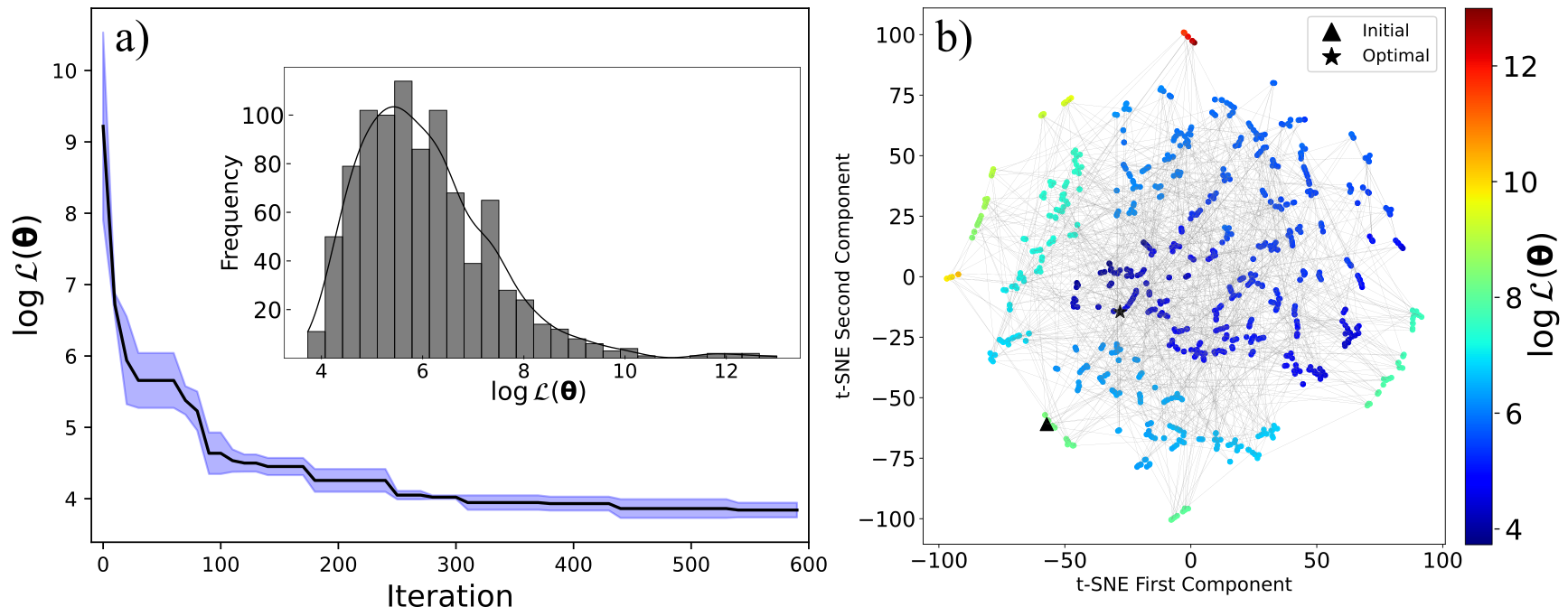}
\caption{(Panel a) Convergence of BO as a function of the iterations. The solid black line and shaded blue region indicate the logarithm of $\Loss(\vtheta)$ (Eq.~\ref{eqn:L_total}) and the standard deviation computed with three independent optimization trajectories. The insight is the histogram of the sampled values of $\log \Loss(\vtheta)$ of one of the BO trajectories. (Panel b) t-SNE visualization of the sampled parameters of a single BO run. Each point represents a sampled parameter set color-coded by its total objective function value. $\blacktriangle$ and $\star$ represent the parameters of the initial and best-found CG model. We used a logarithmic scale to improve the visualization of the ranges for each property's error function.}
\label{fig:BO_learning_curve}
\end{figure}

Optimization of CG models, even with BO, is a dynamic process, as at each iteration BO learns more about $\Loss(\vtheta)$. Fig.~\ref{fig:BO_learning_curve}a illustrates the lowest value of $\Loss$ found by BO as a function of the iterations. To demonstrate BO's robustness, we considered three independent runs, each with an initial random $\vtheta$.
All runs successfully converged to similar optimal solutions with less than 600 iterations. Furthermore, the optimal solutions identified in each run exhibit objective function values several orders of magnitude lower than the initial guesses and by the hybrid strategy. These results suggest that BO effectively avoids regions where $\Loss$ is high (red points in Fig.\ref{fig:BO_learning_curve}b) while concentrating trials in the low areas (blue points in Fig.~\ref{fig:BO_learning_curve}b). 
This is further supported by the histogram of the sampled values of $\Loss$ of a single BO run; insight panel in Fig.~\ref{fig:BO_learning_curve}a. 

It is important to emphasize the scale of this CG model, which consists of 41 parameters. Traditional search methods, such as grid-based approaches, typically require more than 10 trials or grid points per parameter of the CG model. 
Given the complexity of the Pebax CG model, a grid search approach would need more than $10^{40}$ iterations, an infeasible number considering the computational cost of MD simulations to evaluate the three target physical properties. 
Recent studies have also demonstrated the efficiency of BO in optimizing CG models \cite{Weeratunge2023, sestito2020}. However, these previous studies were focused on simpler CG models, leading to lower-dimensional optimization problems. 
To our knowledge, this study is the first to optimize all parameters of a complex CG model, resulting in a high-dimensional optimization problem, 41 dimensions due to the number of parameters in the CG model. The results demonstrate that BO remains effective even for optimization problems with more than a dozen parameters, reinforcing its suitability for complex CG model development.

\begin{figure}[H]
 \centering
 \includegraphics[width=\textwidth]{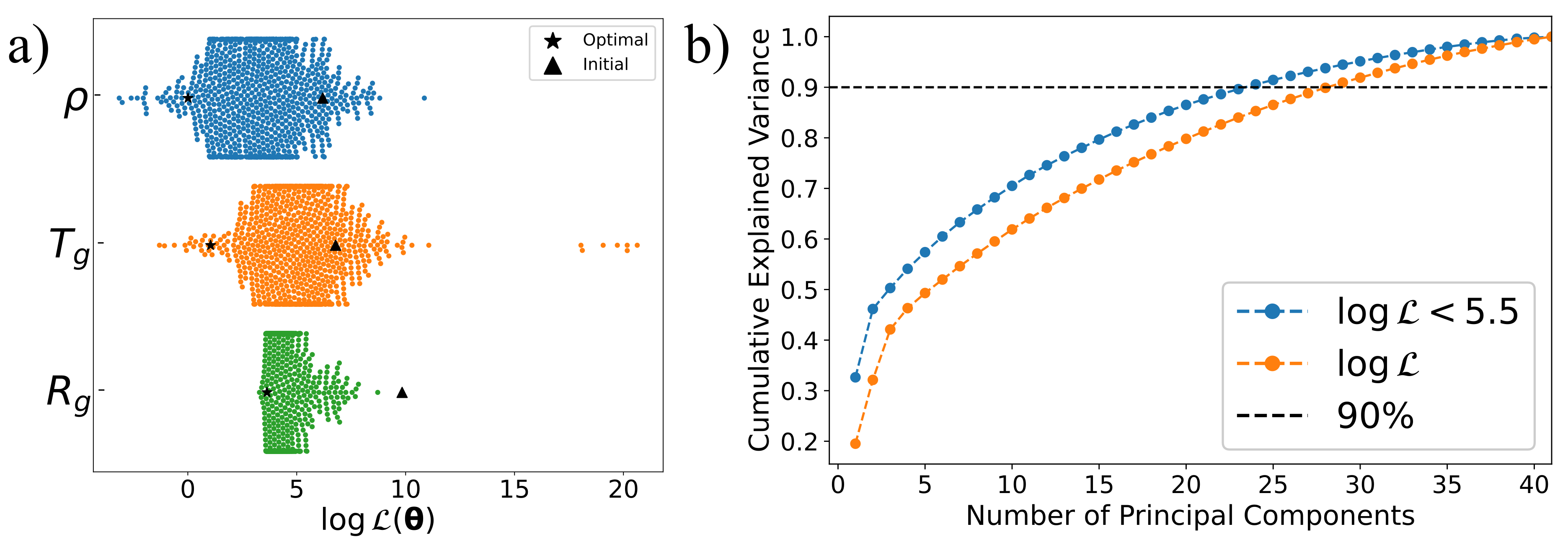}
 \caption{Panel (a) shows the sampled values from BO for each component of the total objective function (Eq.~\ref{eqn:L_total}). 
 A logarithmic scale is used to enhance the visualization of the error functions' range. In panel (a), $\blacktriangle$ and $\star$ mark the parameters of the initial and best models, respectively.
Panel (b) presents the cumulative explained variance as a function of the number of principal components. This variance analysis reflects a similar trend, indicating that the sampling is concentrated in regions where the total loss $\Loss$ is low. 
See the main text for additional details on BO.
}
\label{fig:PCA_loss_by_part}
\end{figure}

To understand why BO is particularly effective for the Pebax CG model, we investigated whether its 41 parameters could be represented in a lower-dimensional space. We first applied Principal Component Analysis (PCA) to the parameters sampled by BO (Fig.~\ref{fig:PCA_loss_by_part}a), finding that 90\% of the variance is captured by the first 28 principal components. While this only reduces the dimensionality by about 32\%, it does indicate some redundancy. To probe for deeper structure, we also performed an ISOMAP analysis, but it did not uncover clear patterns linking low-loss regions, suggesting the absence of a simple non-linear manifold.
Repeating the PCA on samples with $\log \Loss < 5.5$ slightly improved compression: 23 components sufficed to explain 90\% of the variance (Fig.~\ref{fig:PCA_loss_by_part}b). This modest gain suggests that low-loss regions exhibit somewhat more structure. Recent work supports the idea that BO can benefit from linear embeddings in high-dimensional spaces \cite{Priem_2025}. Our results align partially with this view—parameter correlations do emerge in regions of low objective, but retaining 65\% of the original dimensions remains necessary.
Overall, these findings suggest that BO's success here stems less from a highly compressible geometry and more from its ability to adaptively explore a moderately structured, high-dimensional space with some redundancy but no sharply defined low-dimensional manifold.

While the objective function in Eq.~\ref{eqn:L_total} aggregates multiple target properties into a single global function, CG model's optimization can alternatively be approached as a multi-objective optimization problem. In this case, the goal is to identify the Pareto front, which characterizes the trade-offs between competing objectives \cite{rafiei2009,pllana2019,NGUYEN2017}. Although this approach provides a richer understanding of the solution space, it is less practical for high-dimensional CG models as this optimization framework requires a greater number of evaluations. 
Given the linear combination nature of the total objective function, we also analyze the sampled values of each component of $\Loss$. Fig.~\ref{fig:PCA_loss_by_part}a presents the individual values of $\Loss_{\rho}$, $\Loss_{Tg}$, and $\Loss_{Rg}$ sampled by BO.
These results show that BO effectively samples low values ($\log \Loss_i < 5$) for all three components, without a clear preference for any one. Furthermore, these results also indicate that density and glass transition temperature exhibit lower overall relative errors compared to the radius of gyration, suggesting that $Rg$ is the most challenging physical property to reproduce; see Fig.~\ref{fig:density_and_rg}. 
The relatively low errors observed for density and glass transition temperature in some of the different sampled parameters of the CG model (each point in Fig.~\ref{fig:PCA_loss_by_part}b) may be attributed to their intrinsic relationship, as $Tg$ can be determined from the slopes of the density curve; for more details see Ref.~\cite{PATRONE2016}. Overall, our findings suggest that none of the physical properties overwhelmingly dominates the objective function and indicate that BO successfully balances their optimization. This balance enhances the robustness of the resulting coarse-grained models, ensuring a more reliable reproduction of the target physical properties.

\subsection{Physical Properties}
\begin{figure}[H]
\centering
\includegraphics[scale=0.35]{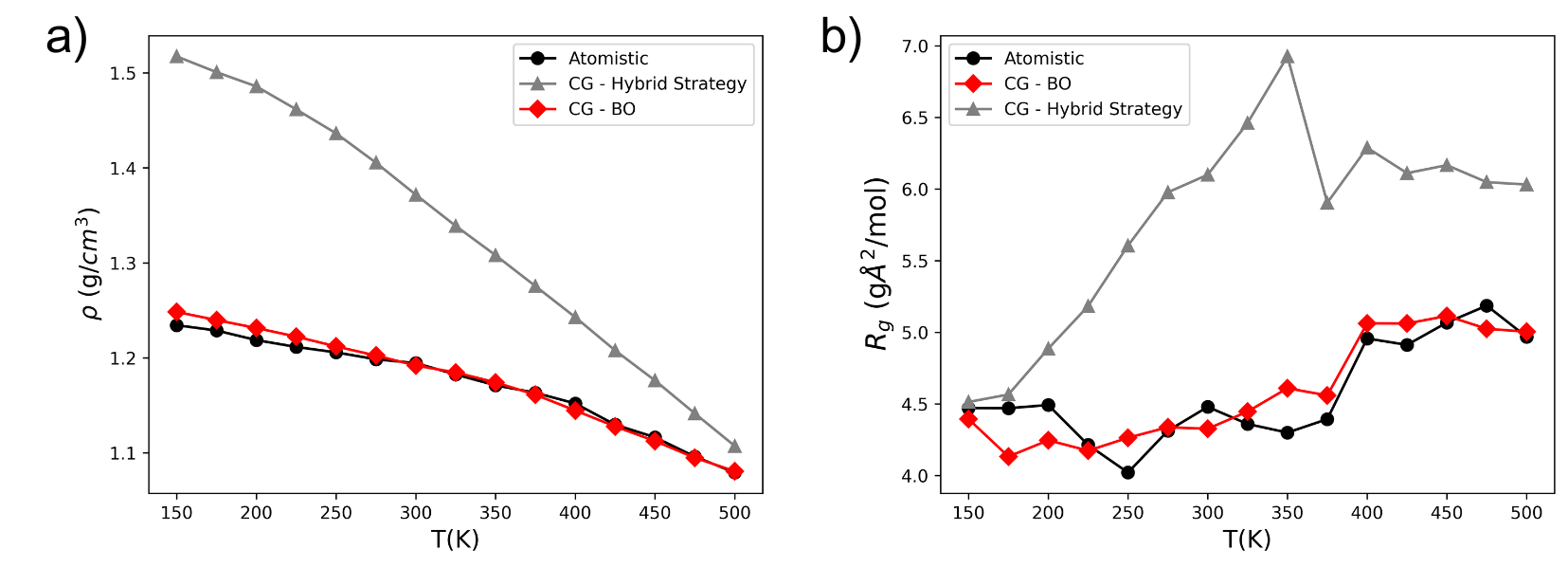}
\caption{Comparison of the density (Panel a) and the radius of gyration (Panel b) for a single Pebax chain using the atomistic model as a reference, alongside the CG-Hybrid Strategy and CG-BO models.}
\label{fig:density_and_rg}
\end{figure}

Fig.~\ref{fig:density_and_rg} compares the density and the radius of gyration for a single Pebax chain, as predicted by three different models: the atomistic reference model, the CG-BO model, and the CG-Hybrid Strategy model, which was developed using a hybrid optimization approach described in the SI. The optimal parameter set $\vtheta$ obtained via BO is reported in the SI, referred to as the CG-BO model. To mitigate the effects of non-linearity in the density profile, only four low-temperature points ($150 \text{K} < T < 225 \text{K}$) and four high-temperature points ($425\;\text{K} < T < 500 \; \text{K}$) were considered in the optimization, as detailed in SI.

Regarding the density (Fig.~\ref{fig:density_and_rg}a), the CG-BO model closely follows the atomistic reference, with only minor deviations at low temperatures ($T < 250$ K). In contrast, the CG-Hybrid Strategy model exhibits significant discrepancies across the entire temperature range, suggesting poor transferability of its parameters, which were originally derived from different polymer systems. It is worth mentioning that the density of the atomistic model deviates from the experimental measurements by 4.3\% \cite{liu2019characterization}. The predicted radius of gyration (Fig.~\ref{fig:density_and_rg}b) further highlights the limitations of the hybrid approach. The CG-Hybrid Strategy model significantly deviates from the atomistic reference, particularly at temperatures above $200$ \text{K}. Conversely, the CG-BO model shows strong agreement with the reference data, effectively capturing the general trend of $Rg$. This demonstrates BO’s capability to systematically explore the parameter space and identify a coarse-grained model that retains key structural characteristics, even within a high-dimensional optimization setting.

For the glass transition temperature ($Tg$), Table \ref{tab:Tg} summarizes the predicted values obtained from the three models. Once again, the CG-BO approach outperforms the CG-Hybrid strategy, producing a significantly lower relative error (10.95\%), while the CG-Hybrid Strategy model exhibits a relative error of 34.62\%.

\begin{table}
\centering
\begin{tblr}{
  cells = {c},
  row{1} = {Mercury},
  hlines,
  vlines,
}
Model                & $Tg$ (K) & $\mathcal{L}_{Tg}(\mathbf{\theta})$\\
Atomistic            & 378.10 &  - \\
CG - BO              & 336.69 &  $0.012$\\
CG - Hybrid Strategy & 247.18 &  $0.119$
\end{tblr}
\caption{The predicted glass transition temperature and its relative error ($\mathcal{L}_{tg}(\mathbf{\theta})$) with the CG-BO and CG-Hybrid Strategy models. We use an atomistic model as the reference.}
\label{tab:Tg}
\end{table}

Lastly, we investigated the sensitivity of the objective function (Eq.~\ref{eqn:L_total}) to changes in the weight parameters $\mathbf{w} = [w_\rho, w_{Rg}, w_{Tg}]$, given our choice of a single-objective optimization approach over a full multi-objective framework. While multi-objective optimization, as proposed by \cite{sestito2020}, is a viable alternative for CG model tuning, it would require significantly more evaluations of $\Loss$ and MD simulations, which becomes impractical for high-dimensional CG models. Furthermore, such an approach would yield a set of Pareto-optimal solutions, making model selection less straightforward.

\begin{figure}
    \centering
    \includegraphics[width=0.8\linewidth]{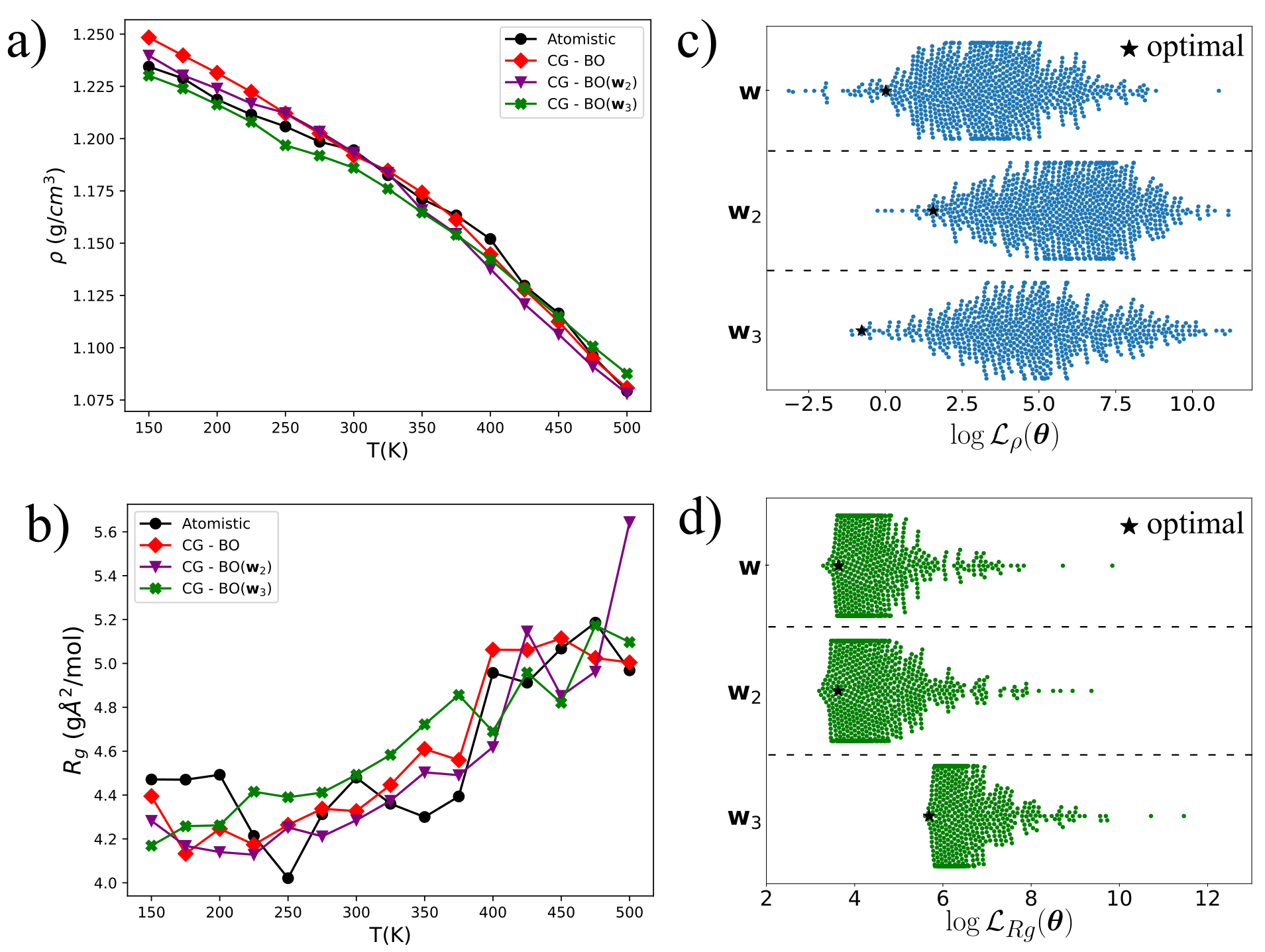}
    \caption{Panels (a) and (b) show the predicted density and radius of gyration, respectively, for the three optimal CG models, CG-BO($\mathbf{w}_i$), obtained using different weight vectors $\mathbf{w}_i$ in the total loss function. Panels (c) and (d) display the sampled values of $\log \Loss_\rho$ and $\log \Loss_{R_g}$ corresponding to each optimization. $\star$ symbols mark the final CG-BO model obtained for each choice of $\mathbf{w}$. Each optimization ran for 600 iterations to ensure convergence. See the main text for further details.}
    \label{fig:w_pannels}
\end{figure}

Fig.~\ref{fig:w_pannels} shows the objective space corresponding to two different weight combinations, $\mathbf{w}_2 = [10, \; 1875, \; 1.5\times10^3]$ and $\mathbf{w}_3 = [5, \; 18750, \; 1.5\times10^3]$, where we deliberately shifted the balance between the three physical properties. 
Each resulting CG model is denoted CG-BO($\mathbf{w}_i$). As seen in Figs.~\ref{fig:w_pannels}a and \ref{fig:w_pannels}b, all three weight configurations lead to similarly accurate predictions for $\rho$ and $Rg$. Fig.~\ref{fig:w_pannels}c and \ref{fig:w_pannels}d show that each $\mathbf{w}_i$ emphasizes a different region of the individual loss components, yet the joint performance remains consistent. The predicted transition temperature in all three optimization cases deviates by $82.94$ K for $\mathbf{w}_2$ and $37.01$ for $\mathbf{w}_3$ from the atomistic model. Overall, the CG-BO($\mathbf{w}_3$) reproduces more accurately $\rho$ and $Tg$, but predicted $Rg$ is $\sim30\%$ less accurate than CG-BO($\mathbf{w}$).
These results indicate that the optimization landscape is relatively robust to moderate changes in $\mathbf{w}$ and that the CG model is flexible enough to capture relevant features regardless of the precise weighting scheme. 


\break
\section{Conclusion}
We demonstrated that Bayesian Optimization enables the efficient optimization of coarse-grained models with a large number of parameters. Unlike traditional optimization methods that often rely on parameter-splitting heuristics to navigate complex search spaces, BO adopts a fundamentally different strategy, using acquisition functions to guide intelligent sampling and globally explore the parameter space. Using BO, we optimized a CG model for the copolymer Pebax without decomposing the problem into smaller sub-tasks, as is typically done in CG model development. Remarkably, BO required fewer than 600 iterations to converge to an optimal set of 41 parameters. The resulting CG model accurately reproduces key physical properties, showing strong agreement with atomistic simulations. It is important to emphasize that, although we used atomistic simulations as the reference in this study, the proposed methodology is more general and can be readily adapted to optimize CG models against experimental data.

This work is motivated by recent advances in extending BO to high-dimensional settings, highlighting its potential beyond traditional low-dimensional parameter tuning. As BO techniques continue to evolve, particularly with the development of trust-region strategies and multi-fidelity frameworks, we anticipate that these tools will further improve the scalability and efficiency of CG model optimization, paving the way for faster, data-driven materials discovery and molecular design.

 
\begin{acknowledgement}
We gratefully acknowledge the support of the RCGI – Research Centre for Greenhouse Gas Innovation, hosted by the University of São Paulo (USP) and sponsored by FAPESP – São Paulo Research Foundation (2014/50279-4 and 2020/15230-5) and Shell Brasil, and the strategic importance of the support given by ANP (Brazil’s National Oil, Natural Gas, and Biofuels Agency) through the R\&D levy regulation. DAD thanks the São Paulo Research Foundation (FAPESP) for financial support through grants 2020/01558-9 and 2022/06973-0. RAVH thanks in full NSERC Discovery Grant No. RGPIN-2024-06594.

\end{acknowledgement}

\bibliography{references}

\providecommand{\latin}[1]{#1}
\makeatletter
\providecommand{\doi}
  {\begingroup\let\do\@makeother\dospecials
  \catcode`\{=1 \catcode`\}=2 \doi@aux}
\providecommand{\doi@aux}[1]{\endgroup\texttt{#1}}
\makeatother
\providecommand*\mcitethebibliography{\thebibliography}
\csname @ifundefined\endcsname{endmcitethebibliography}  {\let\endmcitethebibliography\endthebibliography}{}
\begin{mcitethebibliography}{68}
\providecommand*\natexlab[1]{#1}
\providecommand*\mciteSetBstSublistMode[1]{}
\providecommand*\mciteSetBstMaxWidthForm[2]{}
\providecommand*\mciteBstWouldAddEndPuncttrue
  {\def\EndOfBibitem{\unskip.}}
\providecommand*\mciteBstWouldAddEndPunctfalse
  {\let\EndOfBibitem\relax}
\providecommand*\mciteSetBstMidEndSepPunct[3]{}
\providecommand*\mciteSetBstSublistLabelBeginEnd[3]{}
\providecommand*\EndOfBibitem{}
\mciteSetBstSublistMode{f}
\mciteSetBstMaxWidthForm{subitem}{(\alph{mcitesubitemcount})}
\mciteSetBstSublistLabelBeginEnd
  {\mcitemaxwidthsubitemform\space}
  {\relax}
  {\relax}

\bibitem[Hollingsworth and Dror(2018)Hollingsworth, and Dror]{hollingsworth2018molecular}
Hollingsworth,~S.~A.; Dror,~R.~O. Molecular dynamics simulation for all. \emph{Neuron} \textbf{2018}, \emph{99}, 1129--1143\relax
\mciteBstWouldAddEndPuncttrue
\mciteSetBstMidEndSepPunct{\mcitedefaultmidpunct}
{\mcitedefaultendpunct}{\mcitedefaultseppunct}\relax
\EndOfBibitem
\bibitem[Karplus and McCammon(2002)Karplus, and McCammon]{karplus2002molecular}
Karplus,~M.; McCammon,~J.~A. Molecular dynamics simulations of biomolecules. \emph{Nature structural biology} \textbf{2002}, \emph{9}, 646--652\relax
\mciteBstWouldAddEndPuncttrue
\mciteSetBstMidEndSepPunct{\mcitedefaultmidpunct}
{\mcitedefaultendpunct}{\mcitedefaultseppunct}\relax
\EndOfBibitem
\bibitem[Sadeghi and Noé(2020)Sadeghi, and Noé]{Sadeghi2020}
Sadeghi,~M.; Noé,~F. Large-scale simulation of biomembranes incorporating realistic kinetics into coarse-grained models. \emph{Nature Communications} \textbf{2020}, \emph{11}, 2951\relax
\mciteBstWouldAddEndPuncttrue
\mciteSetBstMidEndSepPunct{\mcitedefaultmidpunct}
{\mcitedefaultendpunct}{\mcitedefaultseppunct}\relax
\EndOfBibitem
\bibitem[Kostritskii and Machtens(2021)Kostritskii, and Machtens]{Kostritskii2021}
Kostritskii,~A.~Y.; Machtens,~J.-P. Molecular mechanisms of ion conduction and ion selectivity in TMEM16 lipid scramblases. \emph{Nature Communications} \textbf{2021}, \emph{12}, 2826\relax
\mciteBstWouldAddEndPuncttrue
\mciteSetBstMidEndSepPunct{\mcitedefaultmidpunct}
{\mcitedefaultendpunct}{\mcitedefaultseppunct}\relax
\EndOfBibitem
\bibitem[Ding \latin{et~al.}(2021)Ding, Lin, and Zhang]{Ding2021}
Ding,~X.; Lin,~X.; Zhang,~B. Stability and folding pathways of tetra-nucleosome from six-dimensional free energy surface. \emph{Nature Communications} \textbf{2021}, \emph{12}, 1091\relax
\mciteBstWouldAddEndPuncttrue
\mciteSetBstMidEndSepPunct{\mcitedefaultmidpunct}
{\mcitedefaultendpunct}{\mcitedefaultseppunct}\relax
\EndOfBibitem
\bibitem[Rahman \latin{et~al.}(2018)Rahman, Lobanova, Jim{\'e}nez-Serratos, Braga, Raptis, M{\"u}ller, Jackson, Avendano, and Galindo]{rahman2018saft}
Rahman,~S.; Lobanova,~O.; Jim{\'e}nez-Serratos,~G.; Braga,~C.; Raptis,~V.; M{\"u}ller,~E.~A.; Jackson,~G.; Avendano,~C.; Galindo,~A. SAFT-$\gamma$ Force Field for the simulation of molecular fluids. 5. Hetero-group coarse-grained models of linear alkanes and the importance of intramolecular interactions. \emph{The Journal of Physical Chemistry B} \textbf{2018}, \emph{122}, 9161--9177\relax
\mciteBstWouldAddEndPuncttrue
\mciteSetBstMidEndSepPunct{\mcitedefaultmidpunct}
{\mcitedefaultendpunct}{\mcitedefaultseppunct}\relax
\EndOfBibitem
\bibitem[Avenda{\~n}o \latin{et~al.}(2011)Avenda{\~n}o, Lafitte, Galindo, Adjiman, Jackson, and M{\"u}ller]{avendano2011saft}
Avenda{\~n}o,~C.; Lafitte,~T.; Galindo,~A.; Adjiman,~C.~S.; Jackson,~G.; M{\"u}ller,~E.~A. SAFT-$\gamma$ force field for the simulation of molecular fluids. 1. A single-site coarse grained model of carbon dioxide. \emph{The Journal of Physical Chemistry B} \textbf{2011}, \emph{115}, 11154--11169\relax
\mciteBstWouldAddEndPuncttrue
\mciteSetBstMidEndSepPunct{\mcitedefaultmidpunct}
{\mcitedefaultendpunct}{\mcitedefaultseppunct}\relax
\EndOfBibitem
\bibitem[Joshi and Deshmukh(2021)Joshi, and Deshmukh]{joshi2021review}
Joshi,~S.~Y.; Deshmukh,~S.~A. A review of advancements in coarse-grained molecular dynamics simulations. \emph{Molecular Simulation} \textbf{2021}, \emph{47}, 786--803\relax
\mciteBstWouldAddEndPuncttrue
\mciteSetBstMidEndSepPunct{\mcitedefaultmidpunct}
{\mcitedefaultendpunct}{\mcitedefaultseppunct}\relax
\EndOfBibitem
\bibitem[Noid(2023)]{noid2023perspective}
Noid,~W.~G. Perspective: Advances, challenges, and insight for predictive coarse-grained models. \emph{The Journal of Physical Chemistry B} \textbf{2023}, \emph{127}, 4174--4207\relax
\mciteBstWouldAddEndPuncttrue
\mciteSetBstMidEndSepPunct{\mcitedefaultmidpunct}
{\mcitedefaultendpunct}{\mcitedefaultseppunct}\relax
\EndOfBibitem
\bibitem[Brini \latin{et~al.}(2013)Brini, Algaer, Ganguly, Li, Rodr{\'\i}guez-Ropero, and van~der Vegt]{brini2013systematic}
Brini,~E.; Algaer,~E.~A.; Ganguly,~P.; Li,~C.; Rodr{\'\i}guez-Ropero,~F.; van~der Vegt,~N.~F. Systematic coarse-graining methods for soft matter simulations--a review. \emph{Soft Matter} \textbf{2013}, \emph{9}, 2108--2119\relax
\mciteBstWouldAddEndPuncttrue
\mciteSetBstMidEndSepPunct{\mcitedefaultmidpunct}
{\mcitedefaultendpunct}{\mcitedefaultseppunct}\relax
\EndOfBibitem
\bibitem[Potter \latin{et~al.}(2019)Potter, Tasche, and Wilson]{potter2019assessing}
Potter,~T.~D.; Tasche,~J.; Wilson,~M.~R. Assessing the transferability of common top-down and bottom-up coarse-grained molecular models for molecular mixtures. \emph{Physical Chemistry Chemical Physics} \textbf{2019}, \emph{21}, 1912--1927\relax
\mciteBstWouldAddEndPuncttrue
\mciteSetBstMidEndSepPunct{\mcitedefaultmidpunct}
{\mcitedefaultendpunct}{\mcitedefaultseppunct}\relax
\EndOfBibitem
\bibitem[Reith \latin{et~al.}(2003)Reith, P{\"u}tz, and M{\"u}ller-Plathe]{reith2003deriving}
Reith,~D.; P{\"u}tz,~M.; M{\"u}ller-Plathe,~F. Deriving effective mesoscale potentials from atomistic simulations. \emph{Journal of computational chemistry} \textbf{2003}, \emph{24}, 1624--1636\relax
\mciteBstWouldAddEndPuncttrue
\mciteSetBstMidEndSepPunct{\mcitedefaultmidpunct}
{\mcitedefaultendpunct}{\mcitedefaultseppunct}\relax
\EndOfBibitem
\bibitem[Izvekov \latin{et~al.}(2004)Izvekov, Parrinello, Burnham, and Voth]{izvekov2004effective}
Izvekov,~S.; Parrinello,~M.; Burnham,~C.~J.; Voth,~G.~A. Effective force fields for condensed phase systems from ab initio molecular dynamics simulation: A new method for force-matching. \emph{The Journal of chemical physics} \textbf{2004}, \emph{120}, 10896--10913\relax
\mciteBstWouldAddEndPuncttrue
\mciteSetBstMidEndSepPunct{\mcitedefaultmidpunct}
{\mcitedefaultendpunct}{\mcitedefaultseppunct}\relax
\EndOfBibitem
\bibitem[Souza \latin{et~al.}(2021)Souza, Alessandri, Barnoud, Thallmair, Faustino, Gr{\"u}newald, Patmanidis, Abdizadeh, Bruininks, Wassenaar, \latin{et~al.} others]{souza2021martini}
Souza,~P.~C.; Alessandri,~R.; Barnoud,~J.; Thallmair,~S.; Faustino,~I.; Gr{\"u}newald,~F.; Patmanidis,~I.; Abdizadeh,~H.; Bruininks,~B.~M.; Wassenaar,~T.~A.; others Martini 3: a general purpose force field for coarse-grained molecular dynamics. \emph{Nature methods} \textbf{2021}, \emph{18}, 382--388\relax
\mciteBstWouldAddEndPuncttrue
\mciteSetBstMidEndSepPunct{\mcitedefaultmidpunct}
{\mcitedefaultendpunct}{\mcitedefaultseppunct}\relax
\EndOfBibitem
\bibitem[Fayaz-Torshizi and M{\"u}ller(2022)Fayaz-Torshizi, and M{\"u}ller]{fayaz2022coarse}
Fayaz-Torshizi,~M.; M{\"u}ller,~E.~A. Coarse-Grained Molecular Simulation of Polymers Supported by the Use of the SAFT-$\gamma$ Mie Equation of State. \emph{Macromolecular Theory and Simulations} \textbf{2022}, \emph{31}, 2100031\relax
\mciteBstWouldAddEndPuncttrue
\mciteSetBstMidEndSepPunct{\mcitedefaultmidpunct}
{\mcitedefaultendpunct}{\mcitedefaultseppunct}\relax
\EndOfBibitem
\bibitem[Avenda{\~n}o \latin{et~al.}(2013)Avenda{\~n}o, Lafitte, Adjiman, Galindo, M{\"u}ller, and Jackson]{avendano2013saft}
Avenda{\~n}o,~C.; Lafitte,~T.; Adjiman,~C.~S.; Galindo,~A.; M{\"u}ller,~E.~A.; Jackson,~G. SAFT-$\gamma$ force field for the simulation of molecular fluids: 2. Coarse-grained models of greenhouse gases, refrigerants, and long alkanes. \emph{The journal of physical chemistry B} \textbf{2013}, \emph{117}, 2717--2733\relax
\mciteBstWouldAddEndPuncttrue
\mciteSetBstMidEndSepPunct{\mcitedefaultmidpunct}
{\mcitedefaultendpunct}{\mcitedefaultseppunct}\relax
\EndOfBibitem
\bibitem[Lobanova \latin{et~al.}(2016)Lobanova, Mejia, Jackson, and M{\"u}eller]{lobanova2016saft}
Lobanova,~O.; Mejia,~A.; Jackson,~G.; M{\"u}eller,~E.~A. SAFT-$\gamma$ force field for the simulation of molecular fluids 6: Binary and ternary mixtures comprising water, carbon dioxide, and n-alkanes. \emph{The Journal of Chemical Thermodynamics} \textbf{2016}, \emph{93}, 320--336\relax
\mciteBstWouldAddEndPuncttrue
\mciteSetBstMidEndSepPunct{\mcitedefaultmidpunct}
{\mcitedefaultendpunct}{\mcitedefaultseppunct}\relax
\EndOfBibitem
\bibitem[Papaioannou \latin{et~al.}(2016)Papaioannou, Calado, Lafitte, Dufal, Sadeqzadeh, Jackson, Adjiman, and Galindo]{papaioannou2016application}
Papaioannou,~V.; Calado,~F.; Lafitte,~T.; Dufal,~S.; Sadeqzadeh,~M.; Jackson,~G.; Adjiman,~C.~S.; Galindo,~A. Application of the SAFT-$\gamma$ Mie group contribution equation of state to fluids of relevance to the oil and gas industry. \emph{Fluid Phase Equilibria} \textbf{2016}, \emph{416}, 104--119\relax
\mciteBstWouldAddEndPuncttrue
\mciteSetBstMidEndSepPunct{\mcitedefaultmidpunct}
{\mcitedefaultendpunct}{\mcitedefaultseppunct}\relax
\EndOfBibitem
\bibitem[Durumeric \latin{et~al.}(2023)Durumeric, Charron, Templeton, Musil, Bonneau, Pasos-Trejo, Chen, Kelkar, No{\'e}, and Clementi]{durumeric2023machine}
Durumeric,~A.~E.; Charron,~N.~E.; Templeton,~C.; Musil,~F.; Bonneau,~K.; Pasos-Trejo,~A.~S.; Chen,~Y.; Kelkar,~A.; No{\'e},~F.; Clementi,~C. Machine learned coarse-grained protein force-fields: Are we there yet? \emph{Current opinion in structural biology} \textbf{2023}, \emph{79}, 102533\relax
\mciteBstWouldAddEndPuncttrue
\mciteSetBstMidEndSepPunct{\mcitedefaultmidpunct}
{\mcitedefaultendpunct}{\mcitedefaultseppunct}\relax
\EndOfBibitem
\bibitem[Shahriari \latin{et~al.}(2016)Shahriari, Swersky, Wang, Adams, and de~Freitas]{BO_review}
Shahriari,~B.; Swersky,~K.; Wang,~Z.; Adams,~R.~P.; de~Freitas,~N. Taking the Human Out of the Loop: A Review of Bayesian Optimization. \emph{Proceedings of the IEEE} \textbf{2016}, \emph{104}, 148--175\relax
\mciteBstWouldAddEndPuncttrue
\mciteSetBstMidEndSepPunct{\mcitedefaultmidpunct}
{\mcitedefaultendpunct}{\mcitedefaultseppunct}\relax
\EndOfBibitem
\bibitem[Ueno \latin{et~al.}(2016)Ueno, Rhone, Hou, Mizoguchi, and Tsuda]{ueno2016combo}
Ueno,~T.; Rhone,~T.~D.; Hou,~Z.; Mizoguchi,~T.; Tsuda,~K. COMBO: An efficient Bayesian optimization library for materials science. \emph{Materials discovery} \textbf{2016}, \emph{4}, 18--21\relax
\mciteBstWouldAddEndPuncttrue
\mciteSetBstMidEndSepPunct{\mcitedefaultmidpunct}
{\mcitedefaultendpunct}{\mcitedefaultseppunct}\relax
\EndOfBibitem
\bibitem[Jalem \latin{et~al.}(2018)Jalem, Kanamori, Takeuchi, Nakayama, Yamasaki, and Saito]{jalem2018bayesian}
Jalem,~R.; Kanamori,~K.; Takeuchi,~I.; Nakayama,~M.; Yamasaki,~H.; Saito,~T. Bayesian-driven first-principles calculations for accelerating exploration of fast ion conductors for rechargeable battery application. \emph{Scientific reports} \textbf{2018}, \emph{8}, 5845\relax
\mciteBstWouldAddEndPuncttrue
\mciteSetBstMidEndSepPunct{\mcitedefaultmidpunct}
{\mcitedefaultendpunct}{\mcitedefaultseppunct}\relax
\EndOfBibitem
\bibitem[Ju \latin{et~al.}(2017)Ju, Shiga, Feng, Hou, Tsuda, and Shiomi]{ju2017designing}
Ju,~S.; Shiga,~T.; Feng,~L.; Hou,~Z.; Tsuda,~K.; Shiomi,~J. Designing nanostructures for phonon transport via Bayesian optimization. \emph{Physical Review X} \textbf{2017}, \emph{7}, 021024\relax
\mciteBstWouldAddEndPuncttrue
\mciteSetBstMidEndSepPunct{\mcitedefaultmidpunct}
{\mcitedefaultendpunct}{\mcitedefaultseppunct}\relax
\EndOfBibitem
\bibitem[Tamura and Hukushima(2018)Tamura, and Hukushima]{tamura2018bayesian}
Tamura,~R.; Hukushima,~K. Bayesian optimization for computationally extensive probability distributions. \emph{Plos one} \textbf{2018}, \emph{13}, e0193785\relax
\mciteBstWouldAddEndPuncttrue
\mciteSetBstMidEndSepPunct{\mcitedefaultmidpunct}
{\mcitedefaultendpunct}{\mcitedefaultseppunct}\relax
\EndOfBibitem
\bibitem[Deng \latin{et~al.}(2020)Deng, Tutunnikov, Averbukh, Thachuk, and Krems]{deng2020bayesian}
Deng,~Z.; Tutunnikov,~I.; Averbukh,~I.~S.; Thachuk,~M.; Krems,~R. Bayesian optimization for inverse problems in time-dependent quantum dynamics. \emph{The Journal of Chemical Physics} \textbf{2020}, \emph{153}\relax
\mciteBstWouldAddEndPuncttrue
\mciteSetBstMidEndSepPunct{\mcitedefaultmidpunct}
{\mcitedefaultendpunct}{\mcitedefaultseppunct}\relax
\EndOfBibitem
\bibitem[Vargas-Hern{\'a}ndez(2020)]{vargas2020}
Vargas-Hern{\'a}ndez,~R.~A. Bayesian Optimization for Calibrating and Selecting Hybrid-Density Functional Models. \emph{The Journal of Physical Chemistry A} \textbf{2020}, \emph{124}, 4053--4061, PMID: 32338905\relax
\mciteBstWouldAddEndPuncttrue
\mciteSetBstMidEndSepPunct{\mcitedefaultmidpunct}
{\mcitedefaultendpunct}{\mcitedefaultseppunct}\relax
\EndOfBibitem
\bibitem[Vargas-Hern{\'a}ndez \latin{et~al.}(2021)Vargas-Hern{\'a}ndez, Chuang, and Brumer]{Vargas2021}
Vargas-Hern{\'a}ndez,~R.~A.; Chuang,~C.; Brumer,~P. {Multi-objective optimization for retinal photoisomerization models with respect to experimental observables}. \emph{The Journal of Chemical Physics} \textbf{2021}, \emph{155}, 234109\relax
\mciteBstWouldAddEndPuncttrue
\mciteSetBstMidEndSepPunct{\mcitedefaultmidpunct}
{\mcitedefaultendpunct}{\mcitedefaultseppunct}\relax
\EndOfBibitem
\bibitem[Singh \latin{et~al.}(2024)Singh, Chuang, and Brumer]{singh2024}
Singh,~D.; Chuang,~C.; Brumer,~P. Machine Learning Optimization of Non-Kasha Behavior and of Transient Dynamics in Model Retinal Isomerization. 2024; \url{https://doi.org/10.1021/acs.jpclett.4c02714}, PMID: 39680656\relax
\mciteBstWouldAddEndPuncttrue
\mciteSetBstMidEndSepPunct{\mcitedefaultmidpunct}
{\mcitedefaultendpunct}{\mcitedefaultseppunct}\relax
\EndOfBibitem
\bibitem[Vargas-Hern{\'a}ndez \latin{et~al.}(2019)Vargas-Hern{\'a}ndez, Guan, Zhang, and Krems]{vargas2019}
Vargas-Hern{\'a}ndez,~R.~A.; Guan,~Y.; Zhang,~D.~H.; Krems,~R.~V. Bayesian optimization for the inverse scattering problem in quantum reaction dynamics. \emph{New Journal of Physics} \textbf{2019}, \emph{21}, 022001\relax
\mciteBstWouldAddEndPuncttrue
\mciteSetBstMidEndSepPunct{\mcitedefaultmidpunct}
{\mcitedefaultendpunct}{\mcitedefaultseppunct}\relax
\EndOfBibitem
\bibitem[Kandasamy \latin{et~al.}(2015)Kandasamy, Schneider, and Poczos]{kandasamy15:HDBO}
Kandasamy,~K.; Schneider,~J.; Poczos,~B. High Dimensional Bayesian Optimisation and Bandits via Additive Models. Proceedings of the 32nd International Conference on Machine Learning. Lille, France, 2015; pp 295--304\relax
\mciteBstWouldAddEndPuncttrue
\mciteSetBstMidEndSepPunct{\mcitedefaultmidpunct}
{\mcitedefaultendpunct}{\mcitedefaultseppunct}\relax
\EndOfBibitem
\bibitem[Wang \latin{et~al.}(2016)Wang, Hutter, Zoghi, Matheson, and De~Freitas]{wang16:HDBO}
Wang,~Z.; Hutter,~F.; Zoghi,~M.; Matheson,~D.; De~Freitas,~N. Bayesian optimization in a billion dimensions via random embeddings. \emph{J. Artif. Int. Res.} \textbf{2016}, \emph{55}, 361–387\relax
\mciteBstWouldAddEndPuncttrue
\mciteSetBstMidEndSepPunct{\mcitedefaultmidpunct}
{\mcitedefaultendpunct}{\mcitedefaultseppunct}\relax
\EndOfBibitem
\bibitem[Nayebi \latin{et~al.}(2019)Nayebi, Munteanu, and Poloczek]{nayebi19a:HDBO}
Nayebi,~A.; Munteanu,~A.; Poloczek,~M. A Framework for {B}ayesian Optimization in Embedded Subspaces. Proceedings of the 36th International Conference on Machine Learning. 2019; pp 4752--4761\relax
\mciteBstWouldAddEndPuncttrue
\mciteSetBstMidEndSepPunct{\mcitedefaultmidpunct}
{\mcitedefaultendpunct}{\mcitedefaultseppunct}\relax
\EndOfBibitem
\bibitem[Eriksson \latin{et~al.}(2019)Eriksson, Pearce, Gardner, Turner, and Poloczek]{eriksson19:HDBO}
Eriksson,~D.; Pearce,~M.; Gardner,~J.; Turner,~R.~D.; Poloczek,~M. Scalable Global Optimization via Local Bayesian Optimization. Advances in Neural Information Processing Systems. 2019\relax
\mciteBstWouldAddEndPuncttrue
\mciteSetBstMidEndSepPunct{\mcitedefaultmidpunct}
{\mcitedefaultendpunct}{\mcitedefaultseppunct}\relax
\EndOfBibitem
\bibitem[Eriksson and Jankowiak(2021)Eriksson, and Jankowiak]{eriksson21a:HDBO}
Eriksson,~D.; Jankowiak,~M. High-dimensional {Bayesian} optimization with sparse axis-aligned subspaces. Proceedings of the Thirty-Seventh Conference on Uncertainty in Artificial Intelligence. 2021; pp 493--503\relax
\mciteBstWouldAddEndPuncttrue
\mciteSetBstMidEndSepPunct{\mcitedefaultmidpunct}
{\mcitedefaultendpunct}{\mcitedefaultseppunct}\relax
\EndOfBibitem
\bibitem[Papenmeier \latin{et~al.}(2022)Papenmeier, Nardi, and Poloczek]{papenmeier22:HDBO}
Papenmeier,~L.; Nardi,~L.; Poloczek,~M. Increasing the Scope as You Learn: Adaptive Bayesian Optimization in Nested Subspaces. Advances in Neural Information Processing Systems. 2022\relax
\mciteBstWouldAddEndPuncttrue
\mciteSetBstMidEndSepPunct{\mcitedefaultmidpunct}
{\mcitedefaultendpunct}{\mcitedefaultseppunct}\relax
\EndOfBibitem
\bibitem[Ziomek and Bou~Ammar(2023)Ziomek, and Bou~Ammar]{ziomek23a:HDBO}
Ziomek,~J.~K.; Bou~Ammar,~H. Are Random Decompositions all we need in High Dimensional {B}ayesian Optimisation? Proceedings of the 40th International Conference on Machine Learning. 2023; pp 43347--43368\relax
\mciteBstWouldAddEndPuncttrue
\mciteSetBstMidEndSepPunct{\mcitedefaultmidpunct}
{\mcitedefaultendpunct}{\mcitedefaultseppunct}\relax
\EndOfBibitem
\bibitem[Hvarfner \latin{et~al.}(2024)Hvarfner, Hellsten, and Nardi]{hvarfner24a:HDBO}
Hvarfner,~C.; Hellsten,~E.~O.; Nardi,~L. Vanilla {B}ayesian Optimization Performs Great in High Dimensions. Proceedings of the 41st International Conference on Machine Learning. 2024; pp 20793--20817\relax
\mciteBstWouldAddEndPuncttrue
\mciteSetBstMidEndSepPunct{\mcitedefaultmidpunct}
{\mcitedefaultendpunct}{\mcitedefaultseppunct}\relax
\EndOfBibitem
\bibitem[Weeratunge \latin{et~al.}(2023)Weeratunge, Robe, and et~al]{Weeratunge2023}
Weeratunge,~H.; Robe,~D.; et~al,~A.~M. {Bayesian coarsening: rapid tuning of polymer model parameters}. \emph{Rheol Acta} \textbf{2023}, 477--490\relax
\mciteBstWouldAddEndPuncttrue
\mciteSetBstMidEndSepPunct{\mcitedefaultmidpunct}
{\mcitedefaultendpunct}{\mcitedefaultseppunct}\relax
\EndOfBibitem
\bibitem[Sestito \latin{et~al.}(2020)Sestito, Thatcher, Shu, Harris, and Wang]{sestito2020}
Sestito,~J.~M.; Thatcher,~M.~L.; Shu,~L.; Harris,~T. A.~L.; Wang,~Y. Coarse-Grained Force Field Calibration Based on Multiobjective Bayesian Optimization to Simulate Water Diffusion in Poly-$\epsilon$-caprolactone. \emph{The Journal of Physical Chemistry A} \textbf{2020}, \emph{124}, 5042--5052, PMID: 32452682\relax
\mciteBstWouldAddEndPuncttrue
\mciteSetBstMidEndSepPunct{\mcitedefaultmidpunct}
{\mcitedefaultendpunct}{\mcitedefaultseppunct}\relax
\EndOfBibitem
\bibitem[Cordina \latin{et~al.}(2023)Cordina, Smith, and Tuttle]{cordina2023}
Cordina,~R.~J.; Smith,~B.; Tuttle,~T. COGITO: A Coarse-Grained Force Field for the Simulation of Macroscopic Properties of Triacylglycerides. \emph{Journal of Chemical Theory and Computation} \textbf{2023}, \emph{19}, 1333--1341, PMID: 36728833\relax
\mciteBstWouldAddEndPuncttrue
\mciteSetBstMidEndSepPunct{\mcitedefaultmidpunct}
{\mcitedefaultendpunct}{\mcitedefaultseppunct}\relax
\EndOfBibitem
\bibitem[Embaye \latin{et~al.}(2021)Embaye, Mart{\'\i}nez-Izquierdo, Malankowska, T{\'e}llez, and Coronas]{embaye2021poly}
Embaye,~A.~S.; Mart{\'\i}nez-Izquierdo,~L.; Malankowska,~M.; T{\'e}llez,~C.; Coronas,~J. Poly (ether-block-amide) copolymer membranes in CO2 separation applications. \emph{Energy \& fuels} \textbf{2021}, \emph{35}, 17085--17102\relax
\mciteBstWouldAddEndPuncttrue
\mciteSetBstMidEndSepPunct{\mcitedefaultmidpunct}
{\mcitedefaultendpunct}{\mcitedefaultseppunct}\relax
\EndOfBibitem
\bibitem[Khalilinejad \latin{et~al.}(2015)Khalilinejad, Sanaeepur, and Kargari]{khalilinejad2015preparation}
Khalilinejad,~I.; Sanaeepur,~H.; Kargari,~A. Preparation of poly (ether-6-block amide)/PVC thin film composite membrane for CO2 separation: effect of top layer thickness and operating parameters. \emph{Journal of Membrane Science and Research} \textbf{2015}, \emph{1}, 124--129\relax
\mciteBstWouldAddEndPuncttrue
\mciteSetBstMidEndSepPunct{\mcitedefaultmidpunct}
{\mcitedefaultendpunct}{\mcitedefaultseppunct}\relax
\EndOfBibitem
\bibitem[Meshkat \latin{et~al.}(2018)Meshkat, Kaliaguine, and Rodrigue]{meshkat2018mixed}
Meshkat,~S.; Kaliaguine,~S.; Rodrigue,~D. Mixed matrix membranes based on amine and non-amine MIL-53 (Al) in Pebax{\textregistered} MH-1657 for CO2 separation. \emph{Separation and Purification Technology} \textbf{2018}, \emph{200}, 177--190\relax
\mciteBstWouldAddEndPuncttrue
\mciteSetBstMidEndSepPunct{\mcitedefaultmidpunct}
{\mcitedefaultendpunct}{\mcitedefaultseppunct}\relax
\EndOfBibitem
\bibitem[Bernardo and Clarizia(2020)Bernardo, and Clarizia]{bernardoenhancing}
Bernardo,~P.; Clarizia,~G. Enhancing Gas Permeation Properties of Pebax{\textregistered} 1657 Membranes via Polysorbate Nonionic Surfactants Doping. \emph{Polymers} \textbf{2020}, \emph{12}\relax
\mciteBstWouldAddEndPuncttrue
\mciteSetBstMidEndSepPunct{\mcitedefaultmidpunct}
{\mcitedefaultendpunct}{\mcitedefaultseppunct}\relax
\EndOfBibitem
\bibitem[Salestan \latin{et~al.}(2021)Salestan, Rahimpour, and Abedini]{salestan2021experimental}
Salestan,~S.~K.; Rahimpour,~A.; Abedini,~R. Experimental and theoretical studies of biopolymers on the efficient CO2/CH4 separation of thin-film Pebax{\textregistered} 1657 membrane. \emph{Chemical Engineering and Processing-Process Intensification} \textbf{2021}, \emph{163}, 108366\relax
\mciteBstWouldAddEndPuncttrue
\mciteSetBstMidEndSepPunct{\mcitedefaultmidpunct}
{\mcitedefaultendpunct}{\mcitedefaultseppunct}\relax
\EndOfBibitem
\bibitem[Liu \latin{et~al.}(2019)Liu, Chen, Lin, Chen, Wu, Lin, and Tung]{liu2019characterization}
Liu,~Y.-C.; Chen,~C.-Y.; Lin,~G.-S.; Chen,~C.-H.; Wu,~K. C.-W.; Lin,~C.-H.; Tung,~K.-L. Characterization and molecular simulation of Pebax-1657-based mixed matrix membranes incorporating MoS2 nanosheets for carbon dioxide capture enhancement. \emph{Journal of Membrane Science} \textbf{2019}, \emph{582}, 358--366\relax
\mciteBstWouldAddEndPuncttrue
\mciteSetBstMidEndSepPunct{\mcitedefaultmidpunct}
{\mcitedefaultendpunct}{\mcitedefaultseppunct}\relax
\EndOfBibitem
\bibitem[Li \latin{et~al.}(2020)Li, Yu, Li, Ma, Zhang, Li, Chang, Zhu, and Xue]{li2020enhanced}
Li,~X.; Yu,~S.; Li,~K.; Ma,~C.; Zhang,~J.; Li,~H.; Chang,~X.; Zhu,~L.; Xue,~Q. Enhanced gas separation performance of Pebax mixed matrix membranes by incorporating ZIF-8 in situ inserted by multiwalled carbon nanotubes. \emph{Separation and Purification Technology} \textbf{2020}, \emph{248}, 117080\relax
\mciteBstWouldAddEndPuncttrue
\mciteSetBstMidEndSepPunct{\mcitedefaultmidpunct}
{\mcitedefaultendpunct}{\mcitedefaultseppunct}\relax
\EndOfBibitem
\bibitem[Xin \latin{et~al.}(2023)Xin, Gao, Ma, Wang, Xuan, Ma, Wei, Zhang, and Zhang]{xin2023preparation}
Xin,~Q.; Gao,~L.; Ma,~F.; Wang,~S.; Xuan,~G.; Ma,~X.; Wei,~M.; Zhang,~L.; Zhang,~Y. Preparation of mixed matrix membrane with high efficiency SO2 separation performance by photosensitive modification and enhanced adsorption of metal--organic framework. \emph{Journal of Materials Science} \textbf{2023}, \emph{58}, 6185--6202\relax
\mciteBstWouldAddEndPuncttrue
\mciteSetBstMidEndSepPunct{\mcitedefaultmidpunct}
{\mcitedefaultendpunct}{\mcitedefaultseppunct}\relax
\EndOfBibitem
\bibitem[Jiang \latin{et~al.}(2021)Jiang, Chuah, Goh, and Wang]{jiang2021facile}
Jiang,~X.; Chuah,~C.~Y.; Goh,~K.; Wang,~R. A facile direct spray-coating of Pebax{\textregistered} 1657: Towards large-scale thin-film composite membranes for efficient CO2/N2 separation. \emph{Journal of Membrane Science} \textbf{2021}, \emph{638}, 119708\relax
\mciteBstWouldAddEndPuncttrue
\mciteSetBstMidEndSepPunct{\mcitedefaultmidpunct}
{\mcitedefaultendpunct}{\mcitedefaultseppunct}\relax
\EndOfBibitem
\bibitem[Materials(2022)]{medea}
Materials,~D. MedeA 3.5 (Materials Exploration and Design Analysis). 2022; \url{www.materialsdesign.com}\relax
\mciteBstWouldAddEndPuncttrue
\mciteSetBstMidEndSepPunct{\mcitedefaultmidpunct}
{\mcitedefaultendpunct}{\mcitedefaultseppunct}\relax
\EndOfBibitem
\bibitem[Sun(1998)]{sun1998}
Sun,~H. COMPASS: An ab Initio Force-Field Optimized for Condensed-Phase Applications Overview with Details on Alkane and Benzene Compounds. \emph{J. Phys. Chem. B} \textbf{1998}, \emph{102}, 7338--7364\relax
\mciteBstWouldAddEndPuncttrue
\mciteSetBstMidEndSepPunct{\mcitedefaultmidpunct}
{\mcitedefaultendpunct}{\mcitedefaultseppunct}\relax
\EndOfBibitem
\bibitem[M{\"u}ller and Jackson(2014)M{\"u}ller, and Jackson]{muller2014force}
M{\"u}ller,~E.~A.; Jackson,~G. Force-field parameters from the SAFT-$\gamma$ equation of state for use in coarse-grained molecular simulations. \emph{Annual review of chemical and biomolecular engineering} \textbf{2014}, \emph{5}, 405--427\relax
\mciteBstWouldAddEndPuncttrue
\mciteSetBstMidEndSepPunct{\mcitedefaultmidpunct}
{\mcitedefaultendpunct}{\mcitedefaultseppunct}\relax
\EndOfBibitem
\bibitem[Mej{\'\i}a \latin{et~al.}(2021)Mej{\'\i}a, M{\"u}ller, and Chaparro~Maldonado]{mejia2021sgtpy}
Mej{\'\i}a,~A.; M{\"u}ller,~E.~A.; Chaparro~Maldonado,~G. SGTPy: A Python code for calculating the interfacial properties of fluids based on the square gradient theory using the {SAFT-VR} Mie equation of state. \emph{Journal of Chemical Information and Modeling} \textbf{2021}, \emph{61}, 1244--1250\relax
\mciteBstWouldAddEndPuncttrue
\mciteSetBstMidEndSepPunct{\mcitedefaultmidpunct}
{\mcitedefaultendpunct}{\mcitedefaultseppunct}\relax
\EndOfBibitem
\bibitem[Ervik \latin{et~al.}(2016)Ervik, Mejia, and M{\"u}ller]{ervik2016bottled}
Ervik,~{\AA}.; Mejia,~A.; M{\"u}ller,~E.~A. Bottled SAFT: A web app providing SAFT-$\gamma$ Mie force field parameters for thousands of molecular fluids. \emph{Journal of chemical information and modeling} \textbf{2016}, \emph{56}, 1609--1614\relax
\mciteBstWouldAddEndPuncttrue
\mciteSetBstMidEndSepPunct{\mcitedefaultmidpunct}
{\mcitedefaultendpunct}{\mcitedefaultseppunct}\relax
\EndOfBibitem
\bibitem[Jim{\'e}nez-Serratos \latin{et~al.}(2017)Jim{\'e}nez-Serratos, Herdes, Haslam, Jackson, and M{\"u}ller]{Guadalupe2017}
Jim{\'e}nez-Serratos,~G.; Herdes,~C.; Haslam,~A.~J.; Jackson,~G.; M{\"u}ller,~E.~A. Group Contribution Coarse-Grained Molecular Simulations of Polystyrene Melts and Polystyrene Solutions in Alkanes Using the SAFT-$\gamma$ Force Field. \emph{Macromolecules} \textbf{2017}, \emph{50}, 4840--4853\relax
\mciteBstWouldAddEndPuncttrue
\mciteSetBstMidEndSepPunct{\mcitedefaultmidpunct}
{\mcitedefaultendpunct}{\mcitedefaultseppunct}\relax
\EndOfBibitem
\bibitem[Richards(2022)]{Richards2022}
Richards,~E. Coarse Grained Models of Surfactants. Ph.D. Thesis, Imperial College London, 2022\relax
\mciteBstWouldAddEndPuncttrue
\mciteSetBstMidEndSepPunct{\mcitedefaultmidpunct}
{\mcitedefaultendpunct}{\mcitedefaultseppunct}\relax
\EndOfBibitem
\bibitem[Lafitte \latin{et~al.}(2012)Lafitte, Avenda{\~n}o, Papaioannou, Galindo, Adjiman, Jackson, and M{\"u}ller]{lafitte2012saft}
Lafitte,~T.; Avenda{\~n}o,~C.; Papaioannou,~V.; Galindo,~A.; Adjiman,~C.~S.; Jackson,~G.; M{\"u}ller,~E.~A. SAFT-$\gamma$ force field for the simulation of molecular fluids: 3. Coarse-grained models of benzene and hetero-group models of n-decylbenzene. \emph{Molecular Physics} \textbf{2012}, \emph{110}, 1189--1203\relax
\mciteBstWouldAddEndPuncttrue
\mciteSetBstMidEndSepPunct{\mcitedefaultmidpunct}
{\mcitedefaultendpunct}{\mcitedefaultseppunct}\relax
\EndOfBibitem
\bibitem[Fayaz-Torshizi and M{\"u}ller(2021)Fayaz-Torshizi, and M{\"u}ller]{fayaz2021}
Fayaz-Torshizi,~M.; M{\"u}ller,~E.~A. Coarse-grained molecular dynamics study of the self-assembly of polyphilic bolaamphiphiles using the SAFT-$\gamma$ Mie force field. \emph{Mol. Syst. Des. Eng.} \textbf{2021}, \emph{6}, 594--608\relax
\mciteBstWouldAddEndPuncttrue
\mciteSetBstMidEndSepPunct{\mcitedefaultmidpunct}
{\mcitedefaultendpunct}{\mcitedefaultseppunct}\relax
\EndOfBibitem
\bibitem[Patrone \latin{et~al.}(2016)Patrone, Dienstfrey, Browning, Tucker, and Christensen]{PATRONE2016}
Patrone,~P.~N.; Dienstfrey,~A.; Browning,~A.~R.; Tucker,~S.; Christensen,~S. Uncertainty quantification in molecular dynamics studies of the glass transition temperature. \emph{Polymer} \textbf{2016}, \emph{87}, 246--259\relax
\mciteBstWouldAddEndPuncttrue
\mciteSetBstMidEndSepPunct{\mcitedefaultmidpunct}
{\mcitedefaultendpunct}{\mcitedefaultseppunct}\relax
\EndOfBibitem
\bibitem[Thompson \latin{et~al.}(2022)Thompson, Aktulga, Berger, Bolintineanu, Brown, Crozier, in~'t Veld, Kohlmeyer, Moore, Nguyen, Shan, Stevens, Tranchida, Trott, and Plimpton]{LAMMPS}
Thompson,~A.~P.; Aktulga,~H.~M.; Berger,~R.; Bolintineanu,~D.~S.; Brown,~W.~M.; Crozier,~P.~S.; in~'t Veld,~P.~J.; Kohlmeyer,~A.; Moore,~S.~G.; Nguyen,~T.~D.; Shan,~R.; Stevens,~M.~J.; Tranchida,~J.; Trott,~C.; Plimpton,~S.~J. {LAMMPS} - a flexible simulation tool for particle-based materials modeling at the atomic, meso, and continuum scales. \emph{Comp. Phys. Comm.} \textbf{2022}, \emph{271}, 108171\relax
\mciteBstWouldAddEndPuncttrue
\mciteSetBstMidEndSepPunct{\mcitedefaultmidpunct}
{\mcitedefaultendpunct}{\mcitedefaultseppunct}\relax
\EndOfBibitem
\bibitem[Watanabe(2023)]{tpe2023}
Watanabe,~S. Tree-structured {P}arzen estimator: Understanding its algorithm components and their roles for better empirical performance. \emph{arXiv preprint arXiv:2304.11127} \textbf{2023}, \relax
\mciteBstWouldAddEndPunctfalse
\mciteSetBstMidEndSepPunct{\mcitedefaultmidpunct}
{}{\mcitedefaultseppunct}\relax
\EndOfBibitem
\bibitem[Garnett(2023)]{bayesoptbook}
Garnett,~R. \emph{{Bayesian Optimization}}; Cambridge University Press, 2023\relax
\mciteBstWouldAddEndPuncttrue
\mciteSetBstMidEndSepPunct{\mcitedefaultmidpunct}
{\mcitedefaultendpunct}{\mcitedefaultseppunct}\relax
\EndOfBibitem
\bibitem[Akiba \latin{et~al.}(2019)Akiba, Sano, Yanase, Ohta, and Koyama]{optuna_2019}
Akiba,~T.; Sano,~S.; Yanase,~T.; Ohta,~T.; Koyama,~M. Optuna: A Next-generation Hyperparameter Optimization Framework. Proceedings of the 25th {ACM} {SIGKDD} International Conference on Knowledge Discovery and Data Mining. 2019\relax
\mciteBstWouldAddEndPuncttrue
\mciteSetBstMidEndSepPunct{\mcitedefaultmidpunct}
{\mcitedefaultendpunct}{\mcitedefaultseppunct}\relax
\EndOfBibitem
\bibitem[Priem \latin{et~al.}(2025)Priem, Diouane, Bartoli, Dubreuil, and Saves]{Priem_2025}
Priem,~R.; Diouane,~Y.; Bartoli,~N.; Dubreuil,~S.; Saves,~P. High-Dimensional Bayesian Optimization Using Both Random and Supervised Embeddings. \emph{AIAA Journal} \textbf{2025}, \emph{63}, 162–173\relax
\mciteBstWouldAddEndPuncttrue
\mciteSetBstMidEndSepPunct{\mcitedefaultmidpunct}
{\mcitedefaultendpunct}{\mcitedefaultseppunct}\relax
\EndOfBibitem
\bibitem[Rafiei \latin{et~al.}(2009)Rafiei, Amirahmadi, and Griva]{rafiei2009}
Rafiei,~S. M.~R.; Amirahmadi,~A.; Griva,~G. Chaos rejection and optimal dynamic response for boost converter using SPEA multi-objective optimization approach. 2009 35th Annual Conference of IEEE Industrial Electronics. 2009; pp 3315--3322\relax
\mciteBstWouldAddEndPuncttrue
\mciteSetBstMidEndSepPunct{\mcitedefaultmidpunct}
{\mcitedefaultendpunct}{\mcitedefaultseppunct}\relax
\EndOfBibitem
\bibitem[Pllana \latin{et~al.}(2019)Pllana, Memeti, and Kolodziej]{pllana2019}
Pllana,~S.; Memeti,~S.; Kolodziej,~J. Customizing Pareto Simulated Annealing for Multi-Objective Optimization of Control Cabinet Layout. 2019 22nd International Conference on Control Systems and Computer Science (CSCS). 2019; pp 78--85\relax
\mciteBstWouldAddEndPuncttrue
\mciteSetBstMidEndSepPunct{\mcitedefaultmidpunct}
{\mcitedefaultendpunct}{\mcitedefaultseppunct}\relax
\EndOfBibitem
\bibitem[Nguyen \latin{et~al.}(2017)Nguyen, {van Iperen}, Raghunath, Abramson, Kipouros, and Somasekharan]{NGUYEN2017}
Nguyen,~H.~A.; {van Iperen},~Z.; Raghunath,~S.; Abramson,~D.; Kipouros,~T.; Somasekharan,~S. Multi-objective optimisation in scientific workflow. \emph{Procedia Computer Science} \textbf{2017}, \emph{108}, 1443--1452, International Conference on Computational Science, ICCS 2017, 12-14 June 2017, Zurich, Switzerland\relax
\mciteBstWouldAddEndPuncttrue
\mciteSetBstMidEndSepPunct{\mcitedefaultmidpunct}
{\mcitedefaultendpunct}{\mcitedefaultseppunct}\relax
\EndOfBibitem
\end{mcitethebibliography}


\providecommand{\latin}[1]{#1}
\makeatletter
\providecommand{\doi}
  {\begingroup\let\do\@makeother\dospecials
  \catcode`\{=1 \catcode`\}=2 \doi@aux}
\providecommand{\doi@aux}[1]{\endgroup\texttt{#1}}
\makeatother
\providecommand*\mcitethebibliography{\thebibliography}
\csname @ifundefined\endcsname{endmcitethebibliography}  {\let\endmcitethebibliography\endthebibliography}{}
\begin{mcitethebibliography}{7}
\providecommand*\natexlab[1]{#1}
\providecommand*\mciteSetBstSublistMode[1]{}
\providecommand*\mciteSetBstMaxWidthForm[2]{}
\providecommand*\mciteBstWouldAddEndPuncttrue
  {\def\EndOfBibitem{\unskip.}}
\providecommand*\mciteBstWouldAddEndPunctfalse
  {\let\EndOfBibitem\relax}
\providecommand*\mciteSetBstMidEndSepPunct[3]{}
\providecommand*\mciteSetBstSublistLabelBeginEnd[3]{}
\providecommand*\EndOfBibitem{}
\mciteSetBstSublistMode{f}
\mciteSetBstMaxWidthForm{subitem}{(\alph{mcitesubitemcount})}
\mciteSetBstSublistLabelBeginEnd
  {\mcitemaxwidthsubitemform\space}
  {\relax}
  {\relax}

\bibitem[Watanabe(2023)]{tpe2023}
Watanabe,~S. Tree-structured {P}arzen estimator: Understanding its algorithm components and their roles for better empirical performance. \emph{arXiv preprint arXiv:2304.11127} \textbf{2023}, \relax
\mciteBstWouldAddEndPunctfalse
\mciteSetBstMidEndSepPunct{\mcitedefaultmidpunct}
{}{\mcitedefaultseppunct}\relax
\EndOfBibitem
\bibitem[Akiba \latin{et~al.}(2019)Akiba, Sano, Yanase, Ohta, and Koyama]{optuna_2019}
Akiba,~T.; Sano,~S.; Yanase,~T.; Ohta,~T.; Koyama,~M. Optuna: A Next-generation Hyperparameter Optimization Framework. Proceedings of the 25th {ACM} {SIGKDD} International Conference on Knowledge Discovery and Data Mining. 2019\relax
\mciteBstWouldAddEndPuncttrue
\mciteSetBstMidEndSepPunct{\mcitedefaultmidpunct}
{\mcitedefaultendpunct}{\mcitedefaultseppunct}\relax
\EndOfBibitem
\bibitem[Patrone \latin{et~al.}(2016)Patrone, Dienstfrey, Browning, Tucker, and Christensen]{PATRONE2016}
Patrone,~P.~N.; Dienstfrey,~A.; Browning,~A.~R.; Tucker,~S.; Christensen,~S. Uncertainty quantification in molecular dynamics studies of the glass transition temperature. \emph{Polymer} \textbf{2016}, \emph{87}, 246--259\relax
\mciteBstWouldAddEndPuncttrue
\mciteSetBstMidEndSepPunct{\mcitedefaultmidpunct}
{\mcitedefaultendpunct}{\mcitedefaultseppunct}\relax
\EndOfBibitem
\bibitem[Halko \latin{et~al.}(2011)Halko, Martinsson, and Tropp]{Halko2011}
Halko,~N.; Martinsson,~P.~G.; Tropp,~J.~A. Finding Structure with Randomness: Probabilistic Algorithms for Constructing Approximate Matrix Decompositions. \emph{SIAM Review} \textbf{2011}, \emph{53}, 217--288\relax
\mciteBstWouldAddEndPuncttrue
\mciteSetBstMidEndSepPunct{\mcitedefaultmidpunct}
{\mcitedefaultendpunct}{\mcitedefaultseppunct}\relax
\EndOfBibitem
\bibitem[Pedregosa \latin{et~al.}(2011)Pedregosa, Varoquaux, Gramfort, Michel, Thirion, Grisel, Blondel, Prettenhofer, Weiss, Dubourg, Vanderplas, Passos, Cournapeau, Brucher, Perrot, and Duchesnay]{scikit-learn}
Pedregosa,~F.; Varoquaux,~G.; Gramfort,~A.; Michel,~V.; Thirion,~B.; Grisel,~O.; Blondel,~M.; Prettenhofer,~P.; Weiss,~R.; Dubourg,~V.; Vanderplas,~J.; Passos,~A.; Cournapeau,~D.; Brucher,~M.; Perrot,~M.; Duchesnay,~E. Scikit-learn: Machine Learning in {P}ython. \emph{Journal of Machine Learning Research} \textbf{2011}, \emph{12}, 2825--2830\relax
\mciteBstWouldAddEndPuncttrue
\mciteSetBstMidEndSepPunct{\mcitedefaultmidpunct}
{\mcitedefaultendpunct}{\mcitedefaultseppunct}\relax
\EndOfBibitem
\bibitem[Van~der Maaten and Hinton(2008)Van~der Maaten, and Hinton]{van2008}
Van~der Maaten,~L.; Hinton,~G. Visualizing data using t-SNE. \emph{Journal of machine learning research} \textbf{2008}, \emph{9}\relax
\mciteBstWouldAddEndPuncttrue
\mciteSetBstMidEndSepPunct{\mcitedefaultmidpunct}
{\mcitedefaultendpunct}{\mcitedefaultseppunct}\relax
\EndOfBibitem
\end{mcitethebibliography}

\end{document}


The purpose of this Supplemental Material is to provide additional details about the proposed work in the main draft. 
The following sections contain all numerical details of the molecular dynamics (MD) simulations and optimization. We also report the values found for the optimal parameters in Section \ref{section:cg-bo_parameters}. 

\section{Numerical details of the Molecular Dynamics Simulations}
\label{section:sm_md}
The central goal of this work is to highlight the use of Bayesian optimization (BO) for the development of CG models. 
The optimization process, as described in the main text, depends on a total error function ($\Loss(\vtheta)$) that quantifies the discrepancy between the physical properties of the predicted target, given a set of parameters ($\vtheta$), and the targets, 
\begin{eqnarray} 
\Loss(\boldsymbol{\theta}) = w_\rho {\cal L}_{\rho}(\boldsymbol{\theta}) + w_{Rg}{\cal L}_{Rg}(\boldsymbol{\theta}) + w_{Tg}{\cal L}_{Tg}(\boldsymbol{\theta}), \label{eqn:L_total} 
\end{eqnarray}
where ${\cal L}_{\rho}$, ${\cal L}_{Rg}$ and ${\cal L}_{Tg}$ are the respective relative errors for the density ($\rho$),  radius of gyration ($Rg$), and the glass transition temperature ($Tg$). The weights, $[w_\rho, w_{R_g}, w_{T_g}]$, in the objective function were chosen to balance the contributions of each term, considering the differences in the number of data points used for their calculations. Specifically, the term associated with $Rg$ was multiplied by $1,875$, a factor derived from the ratio of $15,000$ (the number of points used for density calculation) to 8 (the number of points used for $Rg$). For the glass transition temperature, which depends on a single value, the weight $w_{Tg}$ was set to $15,000$ due to the magnitude of $\Loss_{Tg}$.

Each evaluation of $\Loss$ at the suggested values of $\boldsymbol{\theta}$ by BO follows an iterative loop of molecular dynamics simulations, where the density, radius of gyration, and glass transition temperature are computed. Then the new values collected of $\Loss(\vtheta_i)$ and $\boldsymbol{\theta}_i$ are used to update the Tree-Structured Parzen Estimator (TPE) model \cite{tpe2023}, which will be used in the next iteration of BO to propose a candidate $\vtheta$ which could lead to a lower value in $\Loss$.
All parameters of the CG model were jointly optimized, that is $\boldsymbol{\theta}_i$ includes the bond length ($l$), bond constant ($K_l$), bond angles ($\theta$), angle constants ($k_\theta$), and Mie potential parameters ($\epsilon$, $\sigma$, and $\gamma_r$). 
The TPE model and BO were implemented using the Optuna package \cite{optuna_2019}, with a custom Python program that allows the use of the LAMMPS package for all required MD simulations. To ensure the robustness of the proposed scheme, we performed three independent BO runs, and the optimal parameters found by BO are reported in Tables \ref{tab:optimal_parameters_k_theta}-\ref{tab:optimal_parameters_gamma}.

In the following, we further describe the numerical details in the simulations of the density, glass transition temperature, and radius of gyration.
For density calculation, $\rho$, a simulation box measuring 55 × 55 × 90 $\textup{\AA}$ containing 450 Pebax chains was used. The protocol began with an equilibration phase starting at 500 K, performed using a Langevin integrator, followed by cooling to 300 K and then to 150 K in the NPT ensemble with the Nosé-Hoover integrator. After equilibration, a production phase was conducted, also in the NPT ensemble.

\begin{wrapfigure}{r}{0.4\textwidth}
    \centering
    \includegraphics[width=0.38\textwidth]{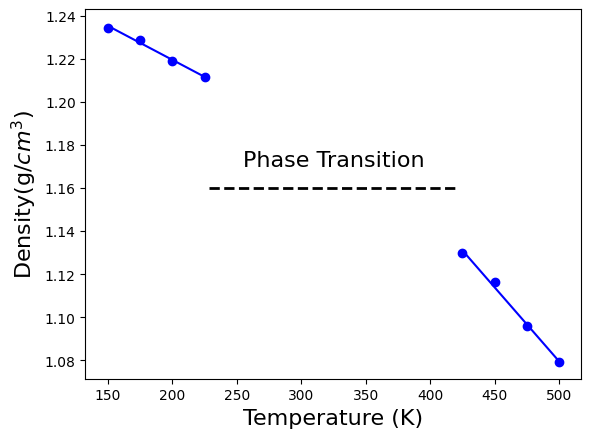}
    \caption{Density values (dots) and the corresponding linear fits used to determine the target property related to density. Data points near the glass transition temperature were excluded due to their non-linear behavior, which would compromise the accuracy of a linear approximation }
    \label{fig:density}
\end{wrapfigure}
At each temperature, $T_i=\{150,175,200,225\}$ and $T_i=\{425,450,475,500\}$ K, density values were measured after 0.3 ns of simulation. The system was gradually heated to reach the next target temperature. A time step of 15 femtoseconds was used. Both time step and simulation durations were determined by monitoring the convergence of density values. The resulting density values were used to fit two linear equations (Fig.~\ref{fig:density}), producing $7,500$ target points for each fit. These points were then used to compute the error between the trial and the atomistic model. 
We also used these fits to obtain the glass temperature, following the procedure described in Ref.~\cite{PATRONE2016}. 
In this procedure, $Tg$ is obtained from the intersection of the two linear fits. 

We obtained the radius of gyration by simulating only a single chain of Pebax in an almost 30 × 30 × 30 $\textup{\AA}$ simulation box. 
The equilibration phase was kept the same, as the procedure used for $\rho$. However, here we evaluated $Rg$ for a wider range of temperatures, starting from 150 K to 500 K, incrementing every 25 K. 
Given the sensitivity of MD simulations of the $Rg$ for a single Pebax, the duration of the chosen timestep, the MD simulations at each temperature were 0.2 ns long with a time step of 0.1 fs.

\section{CG-BO Optimal Parameters}
\label{section:cg-bo_parameters}

Tables S1-S7 show the optimal parameters from both the hybrid strategy and Bayesian Optimization. 

\begin{table}[H]
\centering
\begin{tblr}{
  cells = {c},
  row{1} = {Mercury},
  hlines,
  vlines,
}
$K_\theta$ & Hybrid Strategy (kcal/(mol $rad^2$)) & BO (kcal/(mol $rad^2$))\\
T1-T2-T3  & 2.725 & 5.297  \\
T1-T2-T3  & 6.582 & 7.333 \\
T1-T3-T2  & 4.832 & 11.997 \\
T2-T1-T3  & 5.440 & 6.772 \\
T2-T3-T4  & 3.108 & 11.235 \\
T4-T5-T5  & 3.408 & 6.555 \\
T5-T5-t5  & 2.556 & 1.708 
\end{tblr}
\caption{Optimal parameters for $K_\theta$}
\label{tab:optimal_parameters_k_theta}
\end{table}

\begin{table}[H]
\centering
\begin{tblr}{
  cells = {c},
  row{1} = {Mercury},
  hlines,
  vlines,
}
$\theta$ & Hybrid Strategy (rad) & BO (rad)\\
T1-T2-T3  & 2.393 & 2.134  \\
T1-T2-T3  & 1.782 & 1.715 \\
T1-T3-T2  & 2.347 & 1.512 \\
T2-T1-T3  & 1.890 & 2.898 \\
T2-T3-T4  & 2.353 & 1.279 \\
T4-T5-T5  & 2.383 & 2.276 \\
T5-T5-t5  & 2.389 & 3.012
\end{tblr}
\caption{Optimal parameters for $\theta$}
\label{tab:optimal_parameters_theta}
\end{table}

\begin{table}[H]
\centering
\begin{tblr}{
  cells = {c},
  row{1} = {Mercury},
  hlines,
  vlines,
}
$K_{l}$ & Hybrid Strategy (kcal/(mol$\textup{\AA}$)) & BO (kcal/(mol$\textup{\AA}$))\\
T1-T2  & 6.178 &  4.874 \\
T1-T3  & 4.922 &  4.129 \\
T2-T3  & 10.86 &  9.186 \\
T3-T4  & 7.120 &  6.554 \\
T4-T5  & 5.394 &  5.822 \\
T5-T5  & 9.352 &  1.884
\end{tblr}
\caption{Optimal parameters for  $K_l$}
\label{tab:optimal_parameters_k_l}
\end{table}

\begin{table}[H]
\centering
\begin{tblr}{
  cells = {c},
  row{1} = {Mercury},
  hlines,
  vlines,
}
Bond Length ($\overline{l}$) & Hybrid Strategy ($\textup{\AA}$) & BO ($\textup{\AA}$)\\
T1-T2  & 3.70 & 4.04  \\
T1-T3  & 4.08 & 4.797 \\
T2-T3  & 3.60 & 3.738 \\
T3-T4  & 3.71 & 4.004 \\
T4-T5  & 3.65 & 2.506 \\
T5-T5  & 3.86 & 4.476
\end{tblr}
\caption{Optimal parameters for bond length. For the hybrid strategy, the bond length $\overline{l}_{ij}$ between two sequential beads $i$ and $j$ is set as $\overline{l}_{ij} = (\sigma_i + \sigma_j)/2$.}
\label{tab:optimal_parameters_bond_length}
\end{table}

\begin{table}[H]
\centering
\begin{tblr}{
  cells = {c},
  row{1} = {Mercury},
  hlines,
  vlines,
}
$\epsilon$ & Hybrid Strategy (kcal/mol) & BO (kcal/mol)\\
T1  & 0.673 &  1.453 \\
T2  & 0.749 &  1.440 \\
T3  & 1.262 &  1.391 \\
T4  & 0.507 &  1.169\\
T5  & 0.606 &  0.512
\end{tblr}
\caption{Optimal parameters for  $\epsilon$}
\label{tab:optimal_parameters_epsilon}
\end{table}

\begin{table}[H]
\centering
\begin{tblr}{
  cells = {c},
  row{1} = {Mercury},
  hlines,
  vlines,
}
$\sigma$ & Hybrid Strategy (\textup{\AA}) & BO ($\textup{\AA}$)\\
T1  & 3.220 &  3.940 \\
T2  & 4.180 &  4.553\\
T3  & 3.989 &  2.847\\
T4  & 3.440 &  4.094\\
T5  & 3.860 &  4.773
\end{tblr}
\caption{Optimal parameters for $\sigma$}
\label{tab:tab:optimal_parameters_sigma}
\end{table}

\begin{table}[H]
\centering
\begin{tblr}{
  cells = {c},
  row{1} = {Mercury},
  hlines,
  vlines,
}
$\gamma$ & Hybrid Strategy & BO\\
T1  & 17.850 &  15.837 \\
T2  & 16.400 &  15.629\\
T3  & 34.600 &  10.170 \\
T4  & 12.890 &  11.503 \\
T5  & 12.580 &  11.939
\end{tblr}
\caption{Optimal parameters for $\gamma_r$}
\label{tab:optimal_parameters_gamma}
\end{table}

\section{Effect of $w_\rho$, $w_{Rg}$ and $w_{Tg}$ in $\Loss(\vtheta)$}

\begin{figure}[H]
\includegraphics[width=\textwidth]{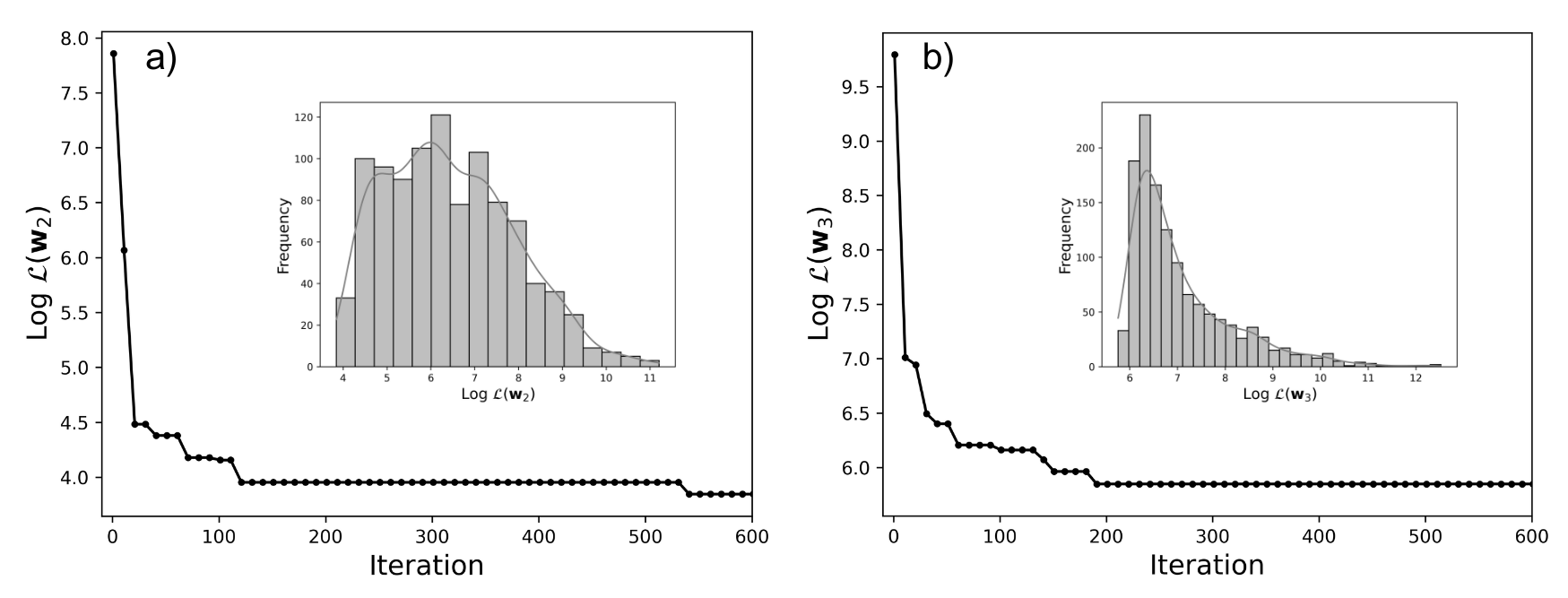}
\caption{Convergence of Bayesian Optimization as a function of the number of iterations. Panel (A) shows the convergence for weight set $\mathbf{w}_2$, while panel (B) corresponds to weight set $\mathbf{w}_3$. In both plots, the black line represents the logarithm of the best objective value, $\log \Loss(\mathbf{w}_i)$ ($i=2,3$), found up to each iteration. The inset in each figure shows a histogram of the sampled values of $\log \Loss(\mathbf{w}_i)$ ($i=2,3$) from the BO run.}\label{fig:BO_learning_curve_weights_SI}
\centering
\end{figure}

To better understand the optimization behavior and its dependence on the weighting scheme, we analyzed the convergence of three independent BO runs using different sets of weights: the original configuration $\mathbf{w} = [w_\rho, w_{Rg}, w_{Tg}]$ with $\mathbf{w} = [1, \; 1875, \;1.5\times10^4]$, and two modified versions, $\mathbf{w}_2 = [10, \; 1875, \; 1.5\times10^3]$ and $\mathbf{w}_3 = [5, \; 18750, \; 1.5\times10^3]$. Fig.~\ref{fig:BO_learning_curve_weights_SI} shows the evolution of the logarithm of the total objective function for the $\mathbf{w}_2$ and $\mathbf{w}_3$ runs. In all three cases, the optimization converged to low-objective regions in fewer than 600 iterations. The inset histograms reveal that although most sampled points span a broad range of objective values, BO consistently targets a narrow region with significantly lower values. This behavior confirms BO’s capacity to efficiently explore and exploit promising areas of the parameter space, even under varying weight configurations.

Beyond convergence, we assessed the physical accuracy of the optimized models by comparing the predicted glass transition temperature ($Tg$) with the atomistic reference. As shown in Table \ref{tab:Tg_si}, all three BO-based models achieved substantially lower relative errors than the CG-Hybrid Strategy. In particular, even when the weights were altered, emphasizing density or downweighting $Tg$, the models optimized for BO maintained a relatively low error in predicting $Tg$. This could be attributed to intrinsic relationships among the physical properties, particularly between density and $Tg$, where the latter is derived from the slope change in the density curve. These findings demonstrate BO's robustness in identifying parameter sets that generalize well across different objective function formulations.

Taken together, these results suggest that the structure of the search space, shaped by physical constraints, enables BO to consistently discover high-quality solutions. The limited sensitivity of the optimization outcomes to the specific weight configuration indicates that precise weight tuning is not a critical factor for BO’s success. This reinforces the potential of Bayesian Optimization as a practical and scalable strategy for high-dimensional coarse-grained (CG) model development, with broader applicability to other complex material and molecular design tasks.

\begin{table}[H]
\centering
\begin{tblr}{
  cells = {c},
  row{1} = {Mercury},
  hlines,
  vlines,
}
Model                               & $Tg$ (K) \\
Atomistic                           & 378.10 \\
CG - BO  ($\mathbf{w}$)             & 336.69 \\
CG - BO ($\mathbf{w_2}$)            & 295.16 \\
CG - BO ($\mathbf{w_3}$)            & 341.09 \\
CG - Hybrid Strategy & 247.18 
\end{tblr}
\caption{Predicted glass transition temperature for the CG-BO model using three different weight sets $\mathbf{w}$, $\mathbf{w}_2$, and $\mathbf{w}_3$ and the CG-Hybrid Strategy. The atomistic model is used as the reference. This comparison highlights the robustness of Bayesian Optimization under different objective weightings.}
\label{tab:Tg_si}
\end{table}

\section{Principal Component Analysis of the Optimal Parameters}
The Principal Component Analysis (PCA) \cite{Halko2011} was performed after the optimization process on the parameter trials. To ensure a balanced contribution of all parameters, they were normalized before the analysis, preventing any single parameter from dominating the variance. Additionally, bond angles were expressed in radians for consistency. The PCA implementation was carried out using the Scikit-learn package \cite{scikit-learn}. 

\section{t-SNE}
The t-distributed Stochastic Neighbor Embedding (t-SNE) analysis \cite{van2008} was performed after the optimization process to visualize the structure of the sampled parameters. The t-SNE projection was carried out using the Scikit-learn package \cite{scikit-learn}.

\bibliography{references}